\documentstyle[preprint,prl,aps,epsf]{revtex}
\begin{document}
\draft

\title{
Quantum Criticality at the Metal Insulator Transition
}

\author{
{\bf D. Schmeltzer}\\
Physics Department, The City College, CUNY\\
Convent Ave. at 138 ST, New York, NY 10031, USA}

\maketitle
\begin{abstract}
We introduce a new method to analysis the many-body problem with disorder.
The method is an extension of the real space renormalization group based
on the operator product expansion. We consider the problem in the presence
of interaction, large elastic mean free path, and finite temperatures. As
a result scaling is stopped either by temperature or the length scale set by
the diverging many-body length scale (superconductivity). Due to disorder
a superconducting instability might take place at $T_{SC}\rightarrow 0$
giving rise to a metallic phase or $T>T_{SC}$. For repulsive interactions
at $T\rightarrow 0$ we flow towards the localized phase which is analized
within the diffusive Finkelstein theory. For finite temperatures with strong
repulsive backward interactions and non-spherical Fermi surfaces characterized
by $|\frac{d\ln N(b)}{\ln b}|\ll 1$ one finds a fixed point $(D^*,\Gamma^*_2)$
in the plane $(D,\Gamma_2^{(s)})$. ($D\propto(K_F\ell)^{-1}$ is the
disorder coupling constant, $\Gamma_2^{(s)}$ is the particle-hole triplet
interaction, $b$ is the length scale and $N(b)$ is the number of channels.)
For weak disorder, $D<D^*$, one obtains a metallic behavior with the resistance
$\rho(D,\Gamma_2^{(s)},T)=\rho(D,\Gamma_2^{(s)},T)\simeq
\rho^*f(\frac{D-D^*}{D^*}\frac{1}{T^{z\nu_1}})$
($\rho^*=\rho(D^*,\Gamma_2^*,1)$, $z=1$, and $\nu_1>1$) in good agreement
with the experiments.\\
\end{abstract}
 
\section{Introduction}
\label{sec-1}

The Metal-Insulator (M-I) transition has been understood within the seminal
paper \cite{01} in 1979. Focusing on noninteracting electrons the authors
demonstrated that in two dimension (2D) even weak disorder is sufficient to
localize the electrons at $T=0$. Few years later \cite{02} it has been
realized by Finkelstein that the particle-hole interaction in the triplet
channel might enhance the conductivity. However a detailed analysis
revealed that at long scale the interaction term diverges making difficult
to determine what will happen at long scales. Recently a remarkable
experiment \cite{03} has been performed on a 2D electron gas in zero
magnetic field strongly points towards a M-I transition in two dimensions.
The characteristic of this experiment performed on a 2DES silicon (
$n_s\sim 10^{11} cm^{-2}$) the mean free path ``$\ell$" is large, the
electron-electron interaction was $\sim 5 mev$, while the Fermi energy is
only $0.6 mev$. The lowest temperature in the experiment was $0.2 K$.
These experimental condition might suggest that the non-linear sigma model
introduced in ref.\cite{02} might not be applicable since it ignores the
interaction effects at length scales shorter than the mean free path.
Since the mean free path is large quantum effects in the momentum range
$2\pi/\ell\leq \mid q\mid \leq \Lambda$ ($\Lambda^{-1}\sim a \sim$
particle separation) might be important for weak disorder,
$\ell\longrightarrow\infty$. This suggests that a phase transition due to a
collective many body interaction might occur before the diffusive limit is
reached. One might have a phase transition from a superconductor to
insulator \cite{04},
Wigner crystal \cite{05,06}, or quantum Hall-insulator transition \cite{07}.
In one dimension it is known that attractive interaction or ferromagnetic
spin fluctuations can suppress the $2k_F$ backscattering leading to a
delocalization transition \cite{08}. We investigate the problem in the
presence of interaction and large mean free paths.
In order to clarify the situation in 2D we propose to use the Renormalization
Group (RG) analysis. Motivated by the fact that the mean free path
``$\ell$" can be large with respect to the particle separation
$a\sim\Lambda^{-1}$ (standard transport theories start at the scale
``$\ell$" and investigate only processes at larger scales governed by
diffusion) we investigate at finite temperatures the competition between
localization and interaction. The competition between multiple scattering
(due to disorder) and the interactions is investigated within a RG theory.
The method used here is different from the procedure used in ref.\cite{02}.
In ref.\cite{02} one emphasizes the disorder by replacing the multiple
elastic scattering by a diffusion theory and in the second step the
interactions are treated perturbatively. We consider a situation where the
elastic mean free path is much larger than any microscopic length.
Therefore we might have a situation that before entering the diffusive
region we have to stop scaling. This can happen if the thermal wave length
is shorter than the elastic mean free path or that the Cooper channel
diverges giving rise to superconductivity.
In the quantum region the single particle
excitations are well-defined and the Fermi surface is parametrized in
terms of $N_o=\frac{\pi k_F}{\Lambda}$ channels. When the cutoff $\Lambda$
is reduced $\Lambda \longrightarrow \Lambda/b$, one finds that the
interactions scale like $\Gamma\longrightarrow \Gamma b^{1-d}$ and the
number of channels, increases like $N=N_ob$ \cite{09}. The disorder scales
like $D\longrightarrow D b^{2-d}$. Due to the fact that the number of
channels increase under scaling, we find that the interaction is marginal
and the disorder is relevant. The quantum region gives rise to a set of
scaling equations for the interaction term $\Gamma$: 
$\Gamma_2^{(c)}$--particle-hole singlet,
$\Gamma_2^{(s)}$--particle-hole triplet, $\Gamma_3^{(s)}$--particle-particle
singlet and disorder $D$ ($d_3^{(s)}$--the Cooperon). Our results show that
due to disorder $\Gamma_3^{(s)}$ might becomes negative resulting in a
superconducting instability at $T\longrightarrow 0$. This might give
rise to an Insulator-Superconductor transition similar to what one has
for superconducting films where a phase transition is
expected \cite{04}. In the absence of an instability the standard method
at length scale $b>b_{Dif}$, $b_{Dif}=\frac{\Lambda}{2\pi/\ell}$ is the
diffusion theory developed by Finkelstein. Here we consider the situation
where the system is in the clean limit such that the microscopic mean free
path $\ell_o=\ell(b=1)$ is large. Due to interaction we obtain that the
mean free path $\ell(b>1)$ increases, $\ell(b)>\ell_o$.\\

In this paper we will
work at finite temperatures such the the thermal wavelength is shorter than
the mean free path $\ell$. We introduce a thermal length scale
$b_T=\frac{v_F\Lambda}{T}$ and consider the situation where $b_{Dif}>b_T$.
Since we have to stop the scaling scaling at $b=b_T$ we are allowed to ignore
the diffusive region. In the recent transport experiment $E_F/T\sim 5$ and
$K_F\ell \gg 5$, therefore the condition $b_{Dif}>b_T$ is realized. The
presence of the cutoff $b_T$ prevent the number of channels to scale to
infinity, instead we have $N_o<N(b)\le N(b_T)=\bar{N}=\frac{E_F}{T}$. We
solve the model under the condition $b_{Dif}>b_T$ and find that the physics
is controlled by the disorder ``$D$" and the particle-hole triplet
$\Gamma_2^{(s)}$. We find that when the number of channels does not scale
(This might be the case at finite temperature or for non-spherical Fermi
surfaces, which obeys $N(b)\simeq \; Const.$), a fixed point in the plane
$\Gamma_2^{(s)}$ and $D$ is obtained. This fixed point separates a metallic
phase from a localized one. The metallic phase is caused by the fact that
the particle-hole triplet flows to a stable fixed point causing a shift in the
critical dimension from $d=2$ to $d<2$. The presence of the stable fixed
point in the triplet channel causes power law behavior of the spin-spin
correlations. The resistivity is expected to obey the scaling behavior:
$\rho(D,\Gamma_2^{(s)},T)=
\rho(D(b),\Gamma_2^{(s)}(b),Tb^z) ; \; \; z\simeq 1$ where
$\Gamma_2^{(s)}(b)=\Gamma_2^*+(\Gamma_2^{(s)}-\Gamma_2^*) b^{-1/\nu_2}$ and
$D(b)=D^*+(D-D^*)b^{1/\nu_1}$. Choosing $Tb^z=T_o$ we obtain:
$\rho(D,\Gamma_2^{(s)},T)\simeq \rho(D^*,\Gamma_2^*,T_o)+
{\textstyle const.}(\frac{D-D^*}{D^*})(\frac{T_o}{T})^{1/z\nu_1}$. In
agreement with the experimental results given in ref.\cite{01} the resistivity
increases for $D>D^*$ and decreases for $D<D^*$.
In the literature alternative theories have been proposed already:
ref.\cite{14} (phenomenological), ref.\cite{10} (within the Finkelstein
theory), as well
as models which focus on the insulating side ref.\cite{15,06}.\\

The plan of this paper is: We introduce in Chapter \ref{sec-2} our microscopic
model. We consider a two dimensional gas in the presence of a screened two-body
potential and a static random potential. We follow a standard method for
treating
disorder. We use the ``replica" method and perform the statistical average
over the disorder. In the second step we parametrize the Fermi Surface (FS)
in terms of $N$ channels. Using this parametrization we identify in Appendix
\ref{app-1} all the possible interaction and disorder terms. We find that the
interaction and disorder is best described in terms of chiral currents
carrying indices of charge, spin, replica, and channel.
In Chapter \ref{sec-3} the method of the Renormalization Group (RG) based on
the Operator Product Expansion (OPE) is introduced. We compute the OPE
rules for the different interaction terms, particle-hole (p-h) singlet,
p-h triplet, particle-particle (p-p) and the Cooperon (the effective
interaction induced by the disorder).
Chapter \ref{sec-4} is devoted to the derivation of the RG equations based
on the OPE results obtained in Chapter \ref{sec-3}.
In Chapter \ref{sec-5} we consider the scaling equations in the quantum limit.
Chapter \ref{sec-6} is devoted to the possible superconducting instability
which might occur in the quantum region.
In Chapter \ref{sec-7} we investigate the scaling equations at finite
temperatures. Here we observe that the physics is determined
by the effective number of channels $\bar{N}$.
In Chapter \ref{sec-8} we solve the RG equations and compute the resistivity.
Chapter \ref{sec-9} is limited to discussions and conclusions.\\

\section{The Microscopic Model}
\label{sec-2}

We introduce the screened two-body potential and perform a statistical average
over the disorder using the replica method. We parametrize the FS in terms
of $N$ Fermions. Using these Fermions we replace the interaction terms and
the Cooperon by chiral currents.
The starting point of our investigation is the averaged disorder \cite{11}
partition function, $\bar{Z^{\alpha}}, \; \alpha=1,...,\alpha\rightarrow 0$,
 
\begin{equation}
\label{eq-001}
  \bar{Z^{\alpha}}=\int D[\bar{\psi},\psi]e^{-S}, \; \; \; \; \;
  \alpha=1,..., \; \alpha\rightarrow 0
\end{equation}

\begin{equation}
\label{eq-002}
  S_o=\int d^dx \int dt \{\sum_{\sigma}\sum_{\alpha}[
\bar{\psi}_{\sigma,\alpha} \partial_t \psi_{\sigma,\alpha}-
\bar{\psi}_{\sigma,\alpha}(\frac{\nabla^2}{2m}+E_F)\psi_{\sigma,\alpha}]\}
\end{equation}
 
\begin{equation}
\label{eq-003}
  S_{int}=\int d^dx\int d^dy\int dt \sum_{\sigma,\sigma^{\prime}}\sum_{\alpha}
\{\bar{\psi}_{\sigma,\alpha}(x)\bar{\psi}_{\sigma^{\prime},\alpha}(y) v(x-y)
\psi_{\sigma^{\prime},\alpha}(y)\psi_{\sigma,\alpha}(x) \}
\end{equation}
 
\begin{equation}
\label{eq-004}
  S_D=-\int dt_1 \int dt_2 \int d^dx\int d^dy \sum_{\sigma,\sigma^{\prime}}
  \sum_{\alpha,\beta} \{\overline{V(x)V(y)}
  \bar{\psi}_{\sigma,\alpha}(x,t_1)\bar{\psi}_{\sigma^{\prime},\beta}(y,t_2)
  \psi_{\sigma^{\prime},\beta}(y,t_2)\psi_{\sigma,\alpha}(x,t_1) \}
\end{equation}
 
\noindent ``$v(x-y)$" is the two body screened potential and
$\overline{V(x)V(y)}=D\delta(x-y)$ where $D=\frac{v_F^2}{K_F\ell}$ is the
disorder parameter controlled by the elastic scattering time $\tau=\ell/v_F$.
Next we parametrize the Fermi surface (FS) in terms of $N$ Fermions or
$N/2$ pairs of right and left movers ( see ref. \cite{09} ):
 
\begin{equation}
\label{eq-005}
  \psi_{\sigma,\alpha}(\vec{x})=\sum_{n=1}^{N/2}
  (e^{ik_F\hat{n}\cdot\vec{x}}R_{n,\sigma,\alpha}(\vec{x})+
  e^{-ik_F\hat{n}\cdot\vec{x}}L_{n,\sigma,\alpha}(\vec{x}))
\end{equation}
 
\noindent $R_{n,\sigma,\alpha}(\vec{x})$ and $L_{n,\sigma,\alpha}(\vec{x})$
are right and left movers defined by momenta $\mid q_{\parallel}\mid<\Lambda$,
$\mid q_{\perp}\mid<\Lambda$ around each Fermi point $k_F=k_F\hat{n}$.
The Fermi momentum is determined by the renormalized Fermi energy
$\bar{E}_F$ which is related to the non-interacting Fermi energy $E_F$ by
the relation $\bar{E}_F=E_F+\delta\mu_F$, such that
$\bar{E}_F=\frac{k_F^2}{2m^*}$. The value of $\delta\mu_F$ is obtained
from the interaction.
The two dimensional Fermions are expressed in terms of the one dimensional
Fermions $\hat{R}_{n,\sigma,\alpha}(x_{\parallel})$ and
$\hat{L}_{n,\sigma,\alpha}(x_{\parallel})$:
 
\[
  R_{n,\sigma,\alpha}(\vec{x})=\hat{R}_{n,\sigma,\alpha}(x_{\parallel})
  Z_n(x_{\perp}), \; \; \; \;\;
  L_{n,\sigma,\alpha}(\vec{x})=\hat{L}_{n,\sigma,\alpha}(x_{\parallel})
  Z_n(x_{\perp})
\]
 
\noindent $Z_n(x_{\perp})$ is scalar function which ensures the conservation
of momentum in the transversal direction. The number of channels (Fermions)
is related to $k_F$ and cutoff $\Lambda<k_F$, $N_o=\frac{\pi k_F}{\Lambda}$.
Using the representation given in Eq.\ref{eq-005}, we introduce the normal
order currents $J^R_{n,\alpha,\sigma}(Z)$ (right mover) and
$J^L_{n,\alpha,\sigma}(\bar{Z})$ (left mover) with $Z$ and $\bar{Z}$
given by $Z=(Z_{\parallel},Z_{\perp})$,
$\bar{Z}=(\bar{Z}_{\parallel},\bar{Z}_{\perp})$,
$Z_{\parallel}=v_Ft-ix_{\parallel}$,
$\bar{Z}_{\parallel}=v_Ft+ix_{\parallel}$,
and $Z_{\perp}=\bar{Z}_{\perp}=x_{\perp}$,

\begin{equation} 
\label{eq-006}
  J^R_{n,\alpha,\sigma}(Z)=:R^{\dagger}_{n,\alpha,\sigma}(Z)
  R_{n,\alpha,\sigma}(Z): \equiv
  R^{\dagger}_{n,\alpha,\sigma}(Z+\epsilon)R_{n,\alpha,\sigma}(Z)-
  \langle R^{\dagger}_{n,\alpha,\sigma}(Z+\epsilon)R_{n,\alpha,\sigma}(Z)
  \rangle_o
\end{equation}

\noindent with $\epsilon=\varepsilon_x-i\delta$, $\epsilon\rightarrow 0$ and
the expectation value:

\begin{equation}
\label{eq-007}
  \langle R^{\dagger}_{n,\alpha,\sigma_1}(\vec{x},t_1)R_{m,\beta,\sigma_2} 
  (\vec{y},t_2)\rangle_o \sim \delta_{n,m} \delta_{\alpha,\beta}
  \delta_{\sigma_1,\sigma_2} \delta_{\Lambda}^{d-1}(x_{\perp}-y_{\perp})
  [v_F(t_1-t_2)-i(x_{\parallel}-y_{\parallel})]^{-1}
\end{equation}

\noindent Similarly we introduce for the left movers:

\begin{equation} 
\label{eq-008}
  J^L_{n,\alpha,\sigma}(\bar{Z})=:L^{\dagger}_{n,\alpha,\sigma}(\bar{Z})
  L_{n,\alpha,\sigma}(\bar{Z}): \equiv
  L^{\dagger}_{n,\alpha,\sigma}(\bar{Z}+\bar{\epsilon})
  L_{n,\alpha,\sigma}(\bar{Z})-\langle
  L^{\dagger}_{n,\alpha,\sigma}(\bar{Z}+\bar{\epsilon}) 
  L_{n,\alpha,\sigma}(\bar{Z}) \rangle_o
\end{equation}

\noindent We write the interaction and the disorder parts in the normal
order form. From the disorder part we obtain the elastic scattering term
$\frac{1}{2\tau} \propto D$ (see ref.\cite{12}). From the disorder part
(Eq.\ref{eq-004}) we obtain the normal order form $\tilde{S}_D$.

From the interaction part we find the normal order representation
$\tilde{S}_{int}$ plus a shift of the Fermi energy: $\delta\mu_{int}
(J^R_{n,\alpha,\sigma}(\vec{x},t)+J^L_{n,\alpha,\sigma}(\vec{x},t))$.
We choose $\delta\mu_F$ such that it cancels the interaction shift,
$\delta\mu_F+\delta\mu_{int}=0$. As a result $S_0$ becomes:

\begin{equation}
\label{eq-009}
  \tilde{S}_o=\sum_{n=1}^{N/2} \sum_{\sigma} \sum_{\alpha} \int d^dx\int dt
\{ \bar{R}_{n,\alpha,\sigma} [\partial_t-v_F\hat{n}\cdot\vec{\partial}]
R_{n,\alpha,\sigma}+
 \bar{L}_{n,\alpha,\sigma} [\partial_t+v_F\hat{n}\cdot\vec{\partial}]
L_{n,\alpha,\sigma}\}
\end{equation}

\noindent Using the representation given in Eq.\ref{eq-005} we replace the
interaction term and disorder in terms of the currents ( see appendix 
\ref{app-1} ).
The interaction part is decomposed in terms of forward
scattering $Q_{n,m}^{(F)}(t,\vec{x},\vec{y})$ (charge part)
$H_{n,m}^{(F)}(t,\vec{x},\vec{y})$ (spin part),
$Q_{n,m}^{(B)}(t,\vec{x},\vec{y})$ (particle-hole in the singlet channel),
$H_{n,m}^{(B)}(t,\vec{x},\vec{y})$ (particle-hole in the triplet channel),
$O_{n,m}^{(s)}(t,\vec{x},\vec{y})$ (particle-particle in the singlet channel),
$O_{n,m}^{(t)}(t,\vec{x},\vec{y})$ (particle-particle in the triplet channel).
From the screened two-body potential $v(\mid\vec{q}\mid)$ we obtain the
scattering matrix elements for the different processes,
$\Gamma^{(c)}(\vec{n},\vec{m})$,$\Gamma^{(s)}(\vec{n},\vec{m})$,
$\Gamma_2^{(c)}(\vec{n},\vec{m})$,$\Gamma_2^{(s)}(\vec{n},\vec{m})$,
$\Gamma_3^{(s)}(\vec{n},\vec{m})$,$\Gamma_3^{(t)}(\vec{n},\vec{m})$.
For the screened case the matrix elements $\Gamma(\vec{n},\vec{m})$
depend only on the angles ``$\theta$" on the FS. For example, if $\kappa$
is the inverse of the screening length we have
$ \Gamma_2^{(s)}(\vec{n},\vec{m})=2\kappa[1+\frac{2k_F}{\kappa}
\cos\theta/2]^{-1}, \; \; 0\leq\theta\leq\pi$
(``$\theta$" is the angle between the unit vectors $\vec{n}$ and $\vec{m}$).
The particle-particle matrix is,
$ \Gamma_3^{(s)}(\vec{n},\vec{m})=\frac{\kappa}{2}[
(1+\frac{2k_F}{\kappa}\sin\theta/2)^{-1}+
(1+\frac{2k_F}{\kappa}\cos\theta/2)^{-1}],\;\; 0\leq\theta\leq\pi$.
We introduce the left and right currents and obtain the representations for
the interaction and disorder terms:
 
\[
  J^R_{n,\alpha,\sigma_1;m,\beta,\sigma_2}(\vec{x},t_1,t_2)=
  :R^{\dagger}_{n,\alpha,\sigma_1}(\vec{x},t_1)
  R_{m,\beta,\sigma_2}(\vec{x},t_2):
\]
 
\begin{equation}   
\label{eq-010}
  J^L_{n,\alpha,\sigma_1;m,\beta,\sigma_2}(\vec{x},t_1,t_2)=
  :L^{\dagger}_{n,\alpha,\sigma_1}(\vec{x},t_1)
  L_{m,\beta,\sigma_2}(\vec{x},t_2):
\end{equation}
 
\noindent For the interaction term we have $t_1=t_2$ and
$\alpha=\beta$. We obtain that the interaction part for a screened two-body
potential takes the form:
 
\[
  \tilde{S}_{int}=\frac{\Lambda^{1-d}}{2N_o}\sum_n\sum_m\int d^dx
  \int dt\sum_{\alpha}
  \{\Gamma^{(c)}(\vec{n},\vec{m})Q^{(F)}_{n,m;\alpha}(\vec{x},t)-
  \Gamma^{(s)}(\vec{n},\vec{m})H^{(F)}_{n,m;\alpha}(\vec{x},t)
\]
 
\begin{equation}
\label{eq-011}
 +\Gamma^{(c)}_2(\vec{n},\vec{m})Q^{(B)}_{n,m;\alpha}(\vec{x},t)-
  \Gamma^{(s)}_2(\vec{n},\vec{m})H^{(B)}_{n,m;\alpha}(\vec{x},t)+
  \Gamma^{(s)}_3(\vec{n},\vec{m})O^{(s)}_{n,m;\alpha}(\vec{x},t)+
  \Gamma^{(t)}_3(\vec{n},\vec{m})O^{(t)}_{n,m;\alpha}(\vec{x},t)\}
\end{equation}

\noindent In Eq.\ref{eq-011} we have to restrict
$\Gamma^{(s)}_3(\vec{n},\vec{m})$ and $\Gamma^{(t)}_3(\vec{n},\vec{m})$
to $\vec{n}\not=\vec{m}$ in order to avoid double counting. If we ignore
the angle dependence of $\Gamma^{(t)}_3$ we have $\Gamma^{(t)}_3\simeq 0$.
For $\vec{n}=\vec{m}$ we have the relation $\Gamma^{(s)}_3(\vec{n},\vec{n})
=\frac{1}{2}\Gamma^{(s)}_2(\vec{n},\vec{n})$. Based on dimensional analysis
we obtain that the interaction term has the dimension of $\Lambda^{1-d}$.
Due to the fact that the interaction is defined at the scale $\Lambda<k_F$
we have the relation
$k_F^{1-d}\Gamma(k_F)=\Lambda^{1-d}(\frac{k_F}{\Lambda})^{d-1}
\Gamma(\Lambda)\propto\frac{\Lambda^{1-d}}{N_o}\Gamma(\Lambda)$
where $N_o=\pi(\frac{k_F}{\Lambda})^{d-1}$ is the number of channels.
The operators $Q^{(F)}_{n,m;\alpha}$, $H^{(F)}_{n,m;\alpha}$,
$Q^{(B)}_{n,m;\alpha}$, $H^{(B)}_{n,m;\alpha}$, $O^{(s)}_{n,m;\alpha}$, and
$O^{(t)}_{n,m;\alpha}$ are given by:
 
\[
  Q^{(F)}_{n,m;\alpha}(\vec{x},t)=J^R_{n,\alpha}(\vec{x},t)
  J^R_{m,\alpha}(\vec{x},t)+J^L_{n,\alpha}(\vec{x},t)J^L_{m,\alpha}(\vec{x},t),
\]
\[
  H^{(F)}_{n,m;\alpha}(\vec{x},t)=\vec{J}^R_{n,\alpha}(\vec{x},t)\cdot
  \vec{J}^R_{m,\alpha}(\vec{x},t)+\vec{J}^L_{n,\alpha}(\vec{x},t)\cdot
  \vec{J}^L_{m,\alpha}(\vec{x},t),
\]
\[
  Q^{(B)}_{n,m;\alpha}(\vec{x},t)=J^R_{n,\alpha}(\vec{x},t)
  J^L_{m,\alpha}(\vec{x},t)+J^L_{n,\alpha}(\vec{x},t)J^R_{m,\alpha}(\vec{x},t),
\]
\[
  H^{(B)}_{n,m;\alpha}(\vec{x},t)=\vec{J}^R_{n,\alpha}(\vec{x},t)\cdot
  \vec{J}^L_{m,\alpha}(\vec{x},t)+\vec{J}^L_{n,\alpha}(\vec{x},t)\cdot
  \vec{J}^R_{m,\alpha}(\vec{x},t),
\]
\[
  O^{(s)}_{n,m;\alpha}(\vec{x},t)=O^{(\parallel)}_{n,m;\alpha}(\vec{x},t)-
  O^{(\perp)}_{n,m;\alpha}(\vec{x},t), \; \;\;
  O^{(t)}_{n,m;\alpha}(\vec{x},t)=O^{(\parallel)}_{n,m;\alpha}(\vec{x},t)+
  O^{(\perp)}_{n,m;\alpha}(\vec{x},t),
\]
\[
  O^{(\perp)}_{n,m;\alpha}(\vec{x},t)=\sum_{\sigma}
  :R^{\dagger}_{n,\alpha,\sigma}(\vec{x},t)R_{m,\alpha,-\sigma}(\vec{x},t):
  :L^{\dagger}_{n,\alpha,-\sigma}(\vec{x},t)L_{m,\alpha,\sigma}(\vec{x},t):
\]
\[
 +:L^{\dagger}_{n,\alpha,\sigma}(\vec{x},t)L_{m,\alpha,-\sigma}(\vec{x},t):
  :R^{\dagger}_{n,\alpha,-\sigma}(\vec{x},t)R_{m,\alpha,\sigma}(\vec{x},t):,
\]
\[
  O^{(\parallel)}_{n,m;\alpha}(\vec{x},t)=\sum_{\sigma}
  :R^{\dagger}_{n,\alpha,\sigma}(\vec{x},t)R_{m,\alpha,\sigma}(\vec{x},t):
  :L^{\dagger}_{n,\alpha,\sigma}(\vec{x},t)L_{m,\alpha,\sigma}(\vec{x},t):
\]
\begin{equation}
\label{eq-012}
 +:L^{\dagger}_{n,\alpha,\sigma}(\vec{x},t)L_{m,\alpha,\sigma}(\vec{x},t):
  :R^{\dagger}_{n,\alpha,\sigma}(\vec{x},t)R_{m,\alpha,\sigma}(\vec{x},t):,
\end{equation}
 
\noindent where
 
\begin{equation}
\label{eq-013}
  J^R_{n,\alpha}(\vec{x},t)=\sum_{\sigma}
  :R^{\dagger}_{n,\alpha,\sigma}(\vec{x},t)
  R_{n,\alpha,\sigma}(\vec{x},t):,\;\;\;\;\;
  \vec{J}^R_{n,\alpha}(\vec{x},t)=\frac{1}{2}
  :R^{\dagger}_{n,\alpha,\sigma_1}(\vec{x},t)
  \vec{\sigma}_{\sigma_1,\sigma_2} R_{n,\alpha,\sigma_2}(\vec{x},t):
\end{equation}

\noindent with similar expressions for the left movers. $``\Lambda"<k_F$
is the cutoff of the theory and we find that the naive dimension of the
interaction field is $\Lambda^{1-d}\;\;(d=2)$. This follows from the fact
that Eq.\ref{eq-009} is invariant under the scaling $\Lambda\longrightarrow
\Lambda/b$, $x=x^{\prime} b$, $t=t^{\prime} b$,
$R_n(\vec{x},t)=b^{-d/2}R_n(\vec{x}^{\prime},t^{\prime})$,
$L_n(\vec{x},t)=b^{-d/2}L_n(\vec{x}^{\prime},t^{\prime})$, and $N(b)=bN_o$.
Following the same procedure as for the interaction we express the disorder
part using again the respective part:
 
\[
  \tilde{S}_D=-\frac{\Lambda^{2-d}}{N_o}\sum_n\sum_m\sum_{\alpha,\beta}
  \int dt_1 \int dt_2 \int d^dx \{
  d^{(d)}_2\rho_{n,m,\alpha,\beta}(\vec{x};t_1,t_2)-
  d^{(c)}_2   q_{n,m,\alpha,\beta}(\vec{x};t_1,t_2)
\]
\begin{equation} 
\label{eq-014}
 -d^{(s)}_2   h^{(B)}_{n,m,\alpha,\beta}(\vec{x};t_1,t_2)+
  d^{(s)}_3   c^{(s)}_{n,m,\alpha,\beta}(\vec{x};t_1,t_2)+
  d^{(t)}_3   c^{(t)}_{n,m,\alpha,\beta}(\vec{x};t_1,t_2)
\end{equation}
 
\noindent The operators in Eq.\ref{eq-014} are in complete analogy with the
ones in Eq.\ref{eq-011}, except
that they are at different times and have double replica index:
 
\[
  \rho_{n,m,\alpha,\beta}(\vec{x};t_1,t_2) \longleftrightarrow
  Q^{(F)}_{n,m,\alpha}(\vec{x},t);
\]
\[
  q_{n,m,\alpha,\beta}(\vec{x};t_1,t_2) \longleftrightarrow
  Q^{(B)}_{n,m,\alpha}(\vec{x},t);
\]
\[
  h^{(B)}_{n,m,\alpha,\beta}(\vec{x};t_1,t_2) \longleftrightarrow
  H^{(B)}_{n,m,\alpha}(\vec{x},t);
\]
\[
  C^{(s)}_{n,m,\alpha,\beta}(\vec{x};t_1,t_2) \longleftrightarrow
  O^{(s)}_{n,m,\alpha}(\vec{x},t);
\]
\[
  C^{(t)}_{n,m,\alpha,\beta}(\vec{x};t_1,t_2) \longleftrightarrow
  O^{(t)}_{n,m,\alpha}(\vec{x},t)
\]
 
\noindent The corresponding constants in Eq.\ref{eq-014} have the initial
values:
$d_3^{(s)}=d_2^{(c)}=\frac{1}{2}d_2^{(s)}\equiv D$, $d_3^{(t)}=0$.
We will find that only the Cooperon term,
$ d^{(s)}_3   c^{(s)}_{n,m,\alpha,\beta}(\vec{x};t_1,t_2)$ is important.
For the rest part of this paper we will ignore the rest of the
terms and consider only the Cooperon part.\\

\section{The Renormalization Group Method}
\label{sec-3}

In the first part of this chapter we will introduce the RG method based on
the OPE. This method is needed in order to analyze the possible phase
diagram of our problem. The real space method based on the Operator
Product Expansion (OPE) introduced in ref.\cite{13} is in particular
advantageous. In order to explain how this works we express the action
in Eq.\ref{eq-011} by a formal expression $S\sim\sum\Gamma_iA_i$
where $A_i$ are the operators and $\Gamma_i$ are the coupling constants. Using
the fact that the time ordered product of the single particle operator is
given by,
\[
  R_{n,\alpha,\sigma}(\vec{x},t_1)
  R^{\dagger}_{m,\beta,\sigma_1}(\vec{y},t_2)\sim
  \frac{1}{2\pi}\delta_{n,m}\delta_{\alpha,\beta}\delta_{\sigma,\sigma_1}
  \delta_{\Lambda}^{d-1}(x_{\perp}-y_{\perp}) \theta(t_1-t_2)
\]
\begin{equation}   
\label{eq-015}
  [v_F(t_1-t_2)-i(x_{\parallel}-y_{\parallel})]^{-1}
\end{equation}

\noindent where $x_{\parallel}=\hat{n}\cdot\vec{x}$,
$x_{\perp}=\vec{x}-\hat{n}\cdot\vec{x}$. We find for any two
operators given in Eq.\ref{eq-011} the OPE:
 
\begin{equation}  
\label{eq-016}
  A_i(\vec{x},t_1)A_j(\vec{x}+a,t_2)\sim\sum_K
   \frac{C^K_{ij}F_K(\mid t_1-t_2\mid)
  A_K(\vec{x},\frac{t_1+t_2}{2})}{[a^2+v_F^2(t_1-t_2)^2]^{x_i+x_j-x_K}}
\end{equation}
 
\noindent with $C^K_{ij}$ the structure constant and
$F_K(\mid t_1-t_2\mid)\sim 1$.
As a result the product of any number of operators can be reduced to a sum
of operators. This implies that once the cutoff $\Lambda$ is reduced to
$\Lambda/b$, one can obtain the scaling equations for coupling constants
$\Gamma_i$. For $\Gamma_i$ with the scaling dimension
$\Gamma_i\longrightarrow\Gamma_ib^{(x_i-d)}$, one obtains:
 
\begin{equation}   
\label{eq-017}
  \frac{d\Gamma_K}{d\ln b}=-(d-x_K)\Gamma_K-\frac{1}{2}
  \sum_{i,j}\tilde{C}_{i,j}^K\Gamma_i\Gamma_j+
  \frac{1}{3!}\sum_{i,j}\sum_{p,q} \tilde{C}_{i,j}^p\tilde{C}_{p,q}^K
  \Gamma_i\Gamma_j\Gamma_q
\end{equation}
 
\noindent where the $\tilde{C}_{i,j}^K$ are proportional to the structure
constants $C_{i,j}^{K}$. In order to be able to complete the RG equation
given in Eq.\ref{eq-017} we have to compute the operator product expansion
of the operators which appear in Eqs.\ref{eq-011} and \ref{eq-014}.
The second part of this chapter will be devoted to the calculation of the
OPE for the interaction and disorder operators. Using current algebra of
the chiral currents given in ref.\cite{16} we will establish the OPE rules
for our problem. The calculation is based on the Wick theorem which replaces
the time order product by the normal ordered form plus all the possible ways
of contracting pairs of Fermion fields. This calculation is standard and
lengthy therefore we will present only the results.
We start with the results for the p-p singlet:

\[
  O^{(s)}_{n,m,\alpha}(\vec{x},t) O^{(s)}_{k,l,\beta}(\vec{x}+\vec{a},t+\tau)=
  \frac{1}{(2\pi)^2}(\frac{\Lambda}{2\pi})^{d-1} \delta_{\alpha,\beta}
  [a^2+(v_F\tau)^2]^{-1} \{ O^{(s)}_{k,m,\alpha}(\vec{x},t)\delta_{n,l}
\]
\begin{equation}  
\label{eq-018}
  + O^{(s)}_{n,l,\alpha}(\vec{x},t)\delta_{k,m}
  -\delta_{n,l}\delta_{k,m}
  (Q^{(B)}_{n,m,\alpha}(\vec{x},t)+Q^{(B)}_{m,n,\alpha}(\vec{x},t))\}
  +[``c" number].
\end{equation}

\noindent From Eq.\ref{eq-018} we learn that the OPE generates p-p and p-h
singlets.\\

For the p-h in the triplet channel no new terms are generated:

\[
  H^{(B)}_{n,m,\alpha}(\vec{x},t) H^{(B)}_{k,l,\beta}(\vec{x}+\vec{a},t+\tau)=
  \frac{1}{(2\pi)^2}(\frac{\Lambda}{2\pi})^{d-1} \delta_{\alpha,\beta}
  [a^2+(v_F\tau)^2]^{-1}
\]
\begin{equation}
\label{eq-019}
  \{ -2H^{(B)}_{n,m,\alpha}(\vec{x},t) [
  \delta_{n,l}\delta_{k,m}+\delta_{n,k}\delta_{l,m}]\} + [``c" number]
\end{equation}

\noindent The p-h singlet generates only a ``c" number:

\begin{equation}
\label{eq-020}
  Q^{(B)}_{n,m,\alpha}(\vec{x},t) Q^{(B)}_{k,l,\beta}(\vec{x}+\vec{a},t+\tau)=
  [``c" number]
\end{equation}

\noindent The OPE for the Cooperon do not generate new terms:

\[
  C^{(s)}_{n,m;\alpha,\beta}(\vec{x};t_1,t_2)
  C^{(s)}_{k,l;\alpha^{\prime}\beta^{\prime}}
  (\vec{x}+\vec{a};t_1+\tau_1,t_2+\tau_2)=
  \frac{1}{(2\pi)^2}(\frac{\Lambda}{2\pi})^{d-1}
  [\frac{1/2}{(v_F\tau_1-ia)(v_F\tau_2+ia)}
\]
\[
  +\frac{1/2}{(v_F\tau_1+ia)(v_F\tau_2-ia)}]
  \{ 2\delta_{\alpha,\beta^{\prime}}\delta_{\alpha^{\prime},\beta}
  [\delta_{m,k}C^{(s)}_{n,l;\alpha,\beta}(\vec{x};t_1,t_2)+
  \delta_{n,l}C^{(s)}_{k,m;\alpha,\beta}(\vec{x};t_1,t_2)]
\]
\begin{equation} 
\label{eq-021}
  -\frac{1}{2}\delta_{\alpha,\alpha^{\prime}}\delta_{\beta,\beta^{\prime}}
  \delta_{m,k}\delta_{n,l} [ C^{(s)}_{n,m;\alpha,\beta}(\vec{x};t_1,t_2)
  +C^{(s)}_{m,n;\alpha,\beta}(\vec{x};t_1,t_2)] \} +[``c" number]
\end{equation}

\noindent The OPE between the p-p and p-h triplet generates the p-p operator
and the p-h singlet:

\[
  O^{(s)}_{n,m;\alpha}(\vec{x},t)H^{(B)}_{k,l;\beta}(\vec{x}+\vec{a},t+\tau)=
  \frac{1}{(2\pi)^2}(\frac{\Lambda}{2\pi})^{d-1}\delta_{\alpha,\beta}
  [a+(v_F\tau)^2]^{-1}\{ \frac{3}{4}\delta_{l,k}\delta_{n,l}\delta_{m,k}
  Q^{(B)}_{n,m})\vec{x},t)]\}
\]
\begin{equation}
\label{eq-022}
  -\frac{\delta_{l,k}}{8}[\delta_{n,l}
  O^{(s)}_{l,m;\alpha}(\vec{x},t)+\delta_{m,l}O^{(s)}_{n,l;\alpha}(\vec{x},t)]
  \}+[``c" number]
\end{equation}

\noindent The OPE between the p-p term and the p-h singlet generates a ``c"
number:

\begin{equation} 
\label{eq-023}
  O^{(B)}_{n,m;\alpha}(\vec{x},t)Q^{(B)}_{k,l;\beta}(\vec{x}+\vec{a},t+\tau)
  =[``c" number]
\end{equation}

\noindent The product for the product p-h triplet and p-h singlet gives a
``c" number:

\begin{equation}  
\label{eq-024}
  H^{(B)}_{n,m;\alpha}(\vec{x},t)Q^{(B)}_{k,l;\beta}(\vec{x}+\vec{a},t+\tau)=
  [``c" number]
\end{equation}

\noindent In the remaining part we present the OPE between the Cooperon and
the interaction operators. For the p-p case we generate the Cooperon and
p-p operator:

\[
  O^{(s)}_{n,m;\gamma}(\vec{x},t)
  C^{(s)}_{k,l;\alpha,\beta}(\vec{x}+\vec{a},t_1+\tau_1,t_2+\tau_2)=
  \frac{1}{(2\pi)^2}(\frac{\Lambda}{2\pi})^{d-1}\{
  [\frac{1/2}{(v_F\tau_1-ia)(v_F\tau_2+ia)}+
\]
\[
  \frac{1/2}{(v_F\tau_1+ia)(v_F\tau_2-ia)}] \delta_{\gamma,\alpha}
  \delta_{\gamma,\beta} [O^{(s)}_{k,m;\gamma}(\vec{x},t)\delta_{n,l}+
  O^{(s)}_{n,l;\gamma}(\vec{x},t)\delta_{k,m}]
  +\frac{1/2}{(v_F\tau_1)^2+a^2}\delta_{k,m}\delta_{n,l}\delta_{n,m}
  (\delta_{\gamma,\alpha}+\delta_{\gamma,\beta})
\]
\begin{equation}
\label{eq-025}
  C^{(s)}_{n,m;\alpha,\beta}(\vec{x},t,t+\tau_2)
  +\frac{1/2}{(v_F\tau_2)^2+a^2}\delta_{k,m}\delta_{n,l}\delta_{n,m}
  (\delta_{\gamma,\alpha}+\delta_{\gamma,\beta})
  C^{(s)}_{n,m;\alpha,\beta}(\vec{x},t+\tau_1,t)\}
  +[``c" number]
\end{equation}

\noindent For the p-h triplet one obtains the p-p and Cooperon terms:

\[
  H^{(B)}_{n,m;\gamma}(\vec{x},t)
  C^{(s)}_{k,l;\alpha,\beta}(\vec{x}+\vec{a},t_1+\tau_1,t_2+\tau_2)=
  \frac{1}{(2\pi)^2}(\frac{\Lambda}{2\pi})^{d-1}\{
  -\frac{1}{2}[\frac{1}{a^2+(v_F\tau_1)^2}
  C^{(s)}_{n,m;\alpha,\beta}(\vec{x},t,t+\tau_2)
\]
\[
  +\frac{1}{a^2+(v_F\tau_2)^2}C^{(s)}_{n,m;\alpha,\beta}(\vec{x},t+\tau_1,t)]
  [\delta_{k,m}\delta_{n,l}\delta_{n,m}(\delta_{\gamma,\alpha}+
  \delta_{\gamma,\beta})\frac{3}{4}]-[\frac{1/2}{(v_F\tau_1-ia)(v_F\tau_2+ia)}
\]
\begin{equation}
\label{eq-026}
  + \frac{1/2}{(v_F\tau_1+ia)(v_F\tau_2-ia)}][\delta_{m,k}\delta_{m,l}+
  \delta_{n,k}\delta_{n,l}][\delta_{\gamma,\alpha}\delta_{\gamma,\beta}]
  [\frac{1}{4}O^{(s)}_{n,m;\gamma}(\vec{x},t)+
  \frac{1}{4}O^{(t)}_{n,m;\gamma}(\vec{x},t)]\}+[``c" number]
\end{equation}

\noindent When we consider the p-h singlet we generate the p-p and Cooperon
terms.

\[ 
  Q^{(B)}_{n,m;\gamma}(\vec{x},t) 
  C^{(s)}_{k,l;\alpha,\beta}(\vec{x}+\vec{a},t_1+\tau_1,t_2+\tau_2)= 
  \frac{1}{(2\pi)^2}(\frac{\Lambda}{2\pi})^{d-1}\{
  -\frac{1}{2}[\frac{1}{a^2+(v_F\tau_1)^2} (
  C^{(s)}_{n,m;\alpha,\beta}(\vec{x},t,t+\tau_2) 
\]
\[
  -C^{(t)}_{n,m;\alpha,\beta}(\vec{x},t,t+\tau_2)]))
  +\frac{1}{a^2+(v_F\tau_2)^2} (C^{(s)}_{n,m;\alpha,\beta}(\vec{x},t+\tau_1,t)
  -C^{(t)}_{n,m;\alpha,\beta}(\vec{x},t+\tau_1,t))]
\]
\[
  \frac{1}{2}(\delta_{n,k}\delta_{m,l}+\delta_{m,k}\delta_{n,l})
  (\delta{\gamma,\alpha}+\delta_{\gamma,\beta})+
  [\frac{1/2}{(v_F\tau_1-ia)(v_F\tau_2+ia)}+
  \frac{1/2}{(v_F\tau_1+ia)(v_F\tau_2-ia)}]
\]
\begin{equation}
\label{eq-027}
  \delta_{\gamma,\alpha}\delta_{\gamma,\beta}
  (\delta_{m,k}\delta_{m,l}+\delta_{n,k}\delta_{n,l})
  (O^{(s)}_{n,m;\gamma}(\vec{x},t)-O^{(t)}_{n,m;\gamma}(\vec{x},t))\}
  +[``c" number]
\end{equation}

\noindent The Eqs.\ref{eq-018}-\ref{eq-027} have been obtained using the
free Fermion action given in Eq.\ref{eq-009}. We will work at a finite
temperature, therefore Eqs.\ref{eq-015} is an approximation of the exact
propagator
$\{\frac{v_F\beta}{\pi}\sin[\frac{\pi}{v_F\beta}(v_F(t_1-t_2)
-i(x_{\parallel}-y_{\parallel}))]\}^{-1}$. This means that the ``$t$"
range of integration in Eqs.\ref{eq-017} and \ref{eq-027} is restricted to
$t<\beta$. \\

\section{Derivation of the RG equations}
\label{sec-4}

This chapter is designated to the computation of the RG equations. This will
be done by expanding the partition function $Z$ in terms of the interaction
and disorder operators. Using the OPE rules derived in
Eqs.\ref{eq-018}-\ref{eq-027} will allow to replace the product of operators
in terms of a sum of operators. When rescaling the minimal distance ``a" to
``ba" will allow to find the scaling equations.
It is important to remark that the
method used here is different from the standard method used for problems
with disorder. The traditional method \cite{02} starts from the diffusion
theory and includes the interaction terms as a perturbation. Here we start
from the Fermion theory and include simultaneously on equal footing the
effects of interaction and disorder. In the standard approach the quantum
diffusion theory ignores completely the effects of interactions at short
distances (distances shorter than the mean free path). We will see that
considering the disorder and interaction on equal footing new terms will
appear in the RG equations. The scaling equations will contain terms which
are controlled by the number of channels.\\

Following the analysis given in section \ref{sec-2} we have:

\begin{equation} 
\label{eq-028}
  \tilde{S}_o=\sum_{n=1}^{N/2} \sum_{\sigma} \sum_{\alpha} \int d^dx\int dt
\{ \bar{R}_{n,\alpha,\sigma} [\partial_t-v_F\hat{n}\cdot\vec{\partial}]
R_{n,\alpha,\sigma}+
\bar{L}_{n,\alpha,\sigma} [\partial_t+v_F\hat{n}\cdot\vec{\partial}]
L_{n,\alpha,\sigma}\}.
\end{equation}

\noindent The action $\tilde{S}_o$ determines the partition function $Z_o$,

\[
  Z_o=\int {\it D}[\bar{\psi},\psi] e^{-\tilde{S}_o}.
\]

\noindent We perturb the partition function $Z_o$ by the interaction
$\tilde{S}_{int}$ and disorder $\tilde{S}_D$:

\[
  \tilde{S}_{int}=\frac{\Lambda^{1-d}}{2N_o}\sum_n\sum_m\sum_{\alpha}
  \int d^dx\int dt \{
  \Gamma^{(c)}(\vec{n},\vec{m})Q^{(F)}_{n,m;\alpha}(\vec{x},t)
  -\Gamma^{(s)}(\vec{n},\vec{m})H^{(F)}_{n,m;\alpha}(\vec{x},t)
  +\Gamma^{(c)}_2(\vec{n},\vec{m})Q^{(B)}_{n,m;\alpha}(\vec{x},t)
\]
\[
  -\Gamma^{(s)}_2(\vec{n},\vec{m})H^{(B)}_{n,m;\alpha}(\vec{x},t)
  +(1-\delta_{n,m})\Gamma^{(s)}_3(\vec{n},\vec{m})
  O^{(s)}_{n,m;\alpha}(\vec{x},t)\}
\]
\[
  =\frac{\Lambda^{1-d}}{2N_o}\sum_n\sum_m\sum_{\alpha}
  \int d^dx\int dt \{
  \Gamma^{(c)}(\vec{n},\vec{m})Q^{(F)}_{n,m;\alpha}(\vec{x},t)
  -\Gamma^{(s)}(\vec{n},\vec{m})H^{(F)}_{n,m;\alpha}(\vec{x},t)  
  +e^{(c)}_2(\vec{n},\vec{m})Q^{(B)}_{n,m;\alpha}(\vec{x},t) 
\] 
\begin{equation} 
\label{eq-029}
  -e^{(s)}_2(\vec{n},\vec{m})H^{(B)}_{n,m;\alpha}(\vec{x},t) 
  +e^{(s)}_3(\vec{n},\vec{m}) O^{(s)}_{n,m;\alpha}(\vec{x},t)\}
\end{equation}

\noindent In Eq.\ref{eq-029} we have ignored for simplicity the
particle-particle triplet and consider only the particle-particle singlet.\\

Due to the relation between the particle-particle and the particle-hole
triplets we remove the term $(1-\delta_{n,m})$ by defining new coupling
constants:

\[
  e^{(c)}_2(\vec{n},\vec{m})=\Gamma^{(c)}_2(\vec{n},\vec{m})-\frac{1}{2}
  \delta_{n,m}\Gamma^{(s)}_3(\vec{n},\vec{m}),
\]
\[
  e^{(s)}_2(\vec{n},\vec{m})=\Gamma^{(s)}_2(\vec{n},\vec{m})-2
  \delta_{n,m}\Gamma^{(s)}_3(\vec{n},\vec{m}),
\]
\begin{equation}   
\label{eq-030}
  e^{(s)}_3(\vec{n},\vec{m})=\Gamma^{(s)}_3(\vec{n},\vec{m}).
\end{equation}

\noindent The results in Eqs.\ref{eq-030} follows from the operator identity

\begin{equation}
\label{eq-031}
  O^{(s)}_{n,m;\alpha}(\vec{x},t)=(1-\delta_{n,m})
  O^{(s)}_{n,m;\alpha}(\vec{x},t)+\delta_{n,m}[\frac{1}{2}
  Q^{(B)}_{n,m;\alpha}(\vec{x},t)-2H^{(B)}_{n,m;\alpha}(\vec{x},t)]
\end{equation}

\noindent Eq.\ref{eq-031} follows directly from the definitions of the
particle-particle singlet for $\vec{n}=\vec{m}$ in terms of the currents
( see Eqs.\ref{eq-012}-\ref{eq-013} ). For $\vec{n}=\vec{m}$ the
particle-hole
triplet $\Gamma^{(s)}_2(\vec{n},\vec{n})$ is related to the particle-particle
singlet $\Gamma^{(s)}_3(\vec{n},\vec{n})$ ( see Eq.\ref{eq-a4} )

\begin{equation}
\label{eq-032}
  \frac{1}{2}\Gamma^{(s)}_2(\vec{n},\vec{n})=\Gamma^{(s)}_3(\vec{n},\vec{n}).
\end{equation}

\noindent For the disorder part we will consider only the dominant
Cooperon term $d^{(s)}_3C^{(s)}_{n,m;\alpha,\beta}(\vec{x};t_1,t_2)$
and we will ignore the effect of forward disorder

\begin{equation} 
\label{eq-033}
  \tilde{S}_D=-\frac{\Lambda^{2-d}}{N_o}\sum_n\sum_m\sum_{\alpha}\sum_{\beta}
  \int d^dx \int dt_1 \int dt_2 \{ d^{(s)}_3
  C^{(s)}_{n,m;\alpha,\beta}(\vec{x};t_1,t_2) \}.
\end{equation}

\noindent Following ref.\cite{13} we compute the partition function $Z$
of the action $\tilde{S}_o+\tilde{S}_{int}+\tilde{S}_D$ by expanding
up to the third order in $\tilde{S}_{int}+\tilde{S}_D$. Using $Z_o$ we obtain:

\[
  Z=Z_o\{1-[\langle\tilde{S}_{int}\rangle_a+\langle\tilde{S}_D\rangle_a
  -\frac{1}{2}\langle\tilde{S}^2_{int}\rangle_a-
  \langle\tilde{S}_{int}\tilde{S}_D\rangle_a-
  \frac{1}{2}\langle\tilde{S}^2_D\rangle_a
\]
\begin{equation} 
\label{eq-034}
  +\frac{1}{3!}\langle\tilde{S}^3_{int}\rangle_a
  +\frac{1}{3!}\langle\tilde{S}^3_D\rangle_a
  +\frac{1}{2}\langle\tilde{S}^2_{int}\tilde{S}_D\rangle_a
  +\frac{1}{2}\langle\tilde{S}_{int}\tilde{S}^2_D\rangle_a ]\}.
\end{equation}

\noindent The meaning of $\langle\cdots\rangle_a$ is to take the expectation
value with respect to $\tilde{S}_o$ defined in Eq.\ref{eq-028}. Since
we want to perform a RG analysis we will take the expectation value only
in the interval $(\Lambda,\Lambda/b)$, $b\geq 1$. In real space this means to
integrate from the microscopic distance $a$ to $ba$. \\

Next we will compute the first term in Eq.\ref{eq-034}

\begin{equation} 
\label{eq-035}
  \langle\tilde{S}_{int}\rangle_{ba}=b^{2-d}\frac{N_o}{N(b)}
  \langle\tilde{S}_{int}\rangle_a, \;\;\;\;\;
  N(b)=N_ob
\end{equation}

\noindent where $\langle\cdots\rangle_{ba}$  represents the expectation
value with respect to Eq.\ref{eq-028} with the new cutoff 
$\Lambda/b=2\pi/ba$.
The expectation value of $\langle\tilde{S}_D\rangle_a$ is different from
Eq.\ref{eq-035}. The difference is due to the two times $t_1$ and $t_2$.
For times $\mid t_1-t_2\mid\leq a/v_F$,
$C^{(s)}_{n,m;\alpha,\beta}(\vec{x};t_1,t_2)$ is replaced by the singlet
particle-particle interaction.

\begin{equation} 
\label{eq-036}
  \langle\tilde{S}_D\rangle_{ba}=b^{3-d}\frac{N_o}{N(b)}
  \langle\tilde{S}_D\rangle_a
\end{equation}
\begin{equation} 
\label{eq-037}
  \Delta\langle\tilde{S}_{int}\rangle_{ba}=
  -\frac{2a}{v_F}b^{2-d}\frac{N_o}{N(b)}\langle\tilde{S}_D(t_1=t_2)\rangle_a.
\end{equation}

\noindent Eq.\ref{eq-037} represents the contribution from the disorder
Cooperon to the singlet particle-particle term when
$\mid t_1-t_2\mid\leq a/v_F$.\\

In order to compute the higher order term we have to use the rule of the
operator product expansion defined in Eqs.\ref{eq-018}-\ref{eq-027}, and
have to perform the time integration. We introduce the notation
$\langle\cdots\rangle_{da}$ which stands for the expectation value in the
domain ($a$,$ba$)
\[
  -\frac{1}{2}\langle\tilde{S}^2_{int}\rangle_{da}=\frac{da}{a}\{
  \frac{\Lambda^{1-d}}{2N}\frac{A^{-1}}{4N}(-1)\sum_n\sum_m\sum_{\alpha}
  \int d^dx\int dt \{ T(\vec{n},\vec{m})Q^{(B)}_{n,m;\alpha}(\vec{x},t)
\]
\begin{equation}
\label{eq-038}
  -S(\vec{n},\vec{m})H^{(B)}_{n,m;\alpha}(\vec{x},t)
  +R(\vec{n},\vec{m})O^{(B)}_{n,m;\alpha}(\vec{x},t)\}
\end{equation}

\noindent where $\frac{da}{a}=\frac{ba-a}{a}\sim d\ln b$. 
$R(\vec{n},\vec{m})$, $S(\vec{n},\vec{m})$, and $T(\vec{n},\vec{m})$ are a
set of polynomials defined by the rules of the OPE given by
Eqs.\ref{eq-018}-\ref{eq-027}.

\[
  T(\vec{n},\vec{m})=-[2(e^{(s)}_3(\vec{n},\vec{m}))^2+\frac{3}{4}\delta_{n,m}
  e^{(s)}_2(\vec{n},\vec{m})e^{(s)}_3(\vec{n},\vec{m})],
\]
\[
  R(\vec{n},\vec{m})=2\sum_{\vec{l}}e^{(s)}_3(\vec{n},\vec{l})
  e^{(s)}_3(\vec{l},\vec{m})+\frac{1}{4}e^{(s)}_3(\vec{n},\vec{m})
  e^{(s)}_2(0),
\]
\begin{equation}  
\label{eq-039}
  S(\vec{n},\vec{m})=4(e^{(s)}_2(\vec{n},\vec{m}))^2
\end{equation}

\noindent where $e^{(s)}_2(0)\equiv e^{(s)}_2(\vec{n},\vec{n})$. $A^{-1}$
is determined by the time integration

\begin{equation}  
\label{eq-040}
  A^{-1}=\frac{I_1(\hat{\beta})}{(2\pi)^{d-1}2\pi v_F};
  \;\;\;\;\;
  \hat{\beta}\equiv\frac{\beta}{\pi a}
\end{equation}

\begin{equation}
\label{eq-041}
  I_1(\hat{\beta})=\frac{2}{\pi}\int_0^{\infty}dx \frac{\cos x/\hat{\beta}}
  {x^2+1}
\end{equation}

\noindent where $\hat{\beta}$ is the dimensionless inverse temperature.
The function $I_1(\hat{\beta})$ originates at $T\not= 0$.
In the limit $\beta\gg 1$,
$I_1(\hat{\beta})\rightarrow 1$. In the limit $\beta\sim 1$,
$I_1(\hat{\beta}) \ll 1$ and the time integration can be neglected.

\begin{equation}  
\label{eq-042}
  -\frac{1}{2}\langle\tilde{S}^2_D\rangle_{da}=\frac{da}{a}\{
  \frac{\Lambda^{2-d}}{N}\frac{B^{-1}}{2N}(-1)\sum_n\sum_m\sum_{\alpha}
  \sum_{\beta}\int d^dx\int dt_1\int dt_2 [2(d^{(s)}_3)^2(1-\frac{1}{8N})N
  C^{(s)}_{n,m;\alpha,\beta}(\vec{x};t_1,t_2)]\}
\end{equation}

\noindent and

\[
  -\langle\tilde{S}_{int}\tilde{S}^2_D\rangle_{da}=\frac{da}{a}\{
  -\frac{\Lambda^{1-d}}{2N}\frac{B^{-1}}{N}\sum_n\sum_m\sum_{\alpha}
  \int d^dx\int dt [-d^{(s)}_3\hat{L}(\vec{n},\vec{m})
  O^{(s)}_{n,m;\alpha}(\vec{x},t)]
\]
\begin{equation}
\label{eq-043}
  -\frac{\Lambda^{2-d}}{N}\frac{A^{-1}}{2N}\sum_n\sum_m\sum_{\alpha}
  \sum_{\beta}
  \int d^dx\int dt_1\int dt_2 [-d^{(s)}_3\hat{M}(\vec{n},\vec{m}) 
  C^{(s)}_{n,m;\alpha,\beta}(\vec{x};t_1,t_2)]\}
\end{equation}

\noindent where

\begin{equation}    
\label{eq-044}
  B^{-1}=\frac{I_2(2\hat{\beta})}{(2\pi)^{d-1}2v_F^2}
\end{equation}

\begin{equation}     
\label{eq-045}
  I_2(\hat{\beta})=[I_1(\hat{\beta})]^2
\end{equation}

\noindent The term $\hat{L}(\vec{n},\vec{m})$ and $\hat{M}(\vec{n},\vec{m})$
are given by:

\begin{equation}      
\label{eq-046}
  \hat{L}(\vec{n},\vec{m})=2\sum_{\vec{l}}e^{(s)}_3(\vec{l},\vec{m})
  +\frac{1}{2}e^{(s)}_2(\vec{n},\vec{m})-2e^{(c)}_2(\vec{n},\vec{m})
\end{equation}

\noindent and

\begin{equation}       
\label{eq-047}
  \hat{M}(\vec{n},\vec{m})=\frac{3}{2}e^{(s)}_2(\vec{n},\vec{m})
  +2\gamma^{(s)}_3(\vec{n},\vec{m})\delta_{n,m}-2e^{(c)}_2(\vec{n},\vec{m})
\end{equation} 

\noindent The set of Eqs.\ref{eq-043}-\ref{eq-047} concludes the RG 
calculation to second order.\\

The presence of the elastic mean free path introduces a cutoff in the time
domain and allows us to apply the method of OPE to higher order. To third
order in the interaction parameters $e^{(s)}_2$, $e^{(c)}_2$, $e^{(s)}_3$,
and disorder $d^{(s)}_3$ we obtain:

\[
  \langle\tilde{S}^2_{int}\tilde{S}_D\rangle_{da}=\frac{da}{a}\{
  \frac{\Lambda^{2-d}}{N}\frac{A^{-1}}{2N^2}\sum_n\sum_m\sum_{\alpha}
  \sum_{\beta} \int d^dx\int dt_1\int dt_2 [ \frac{A^{-1}}{2}
  \frac{J_1(\hat{\beta})}{I_1(\hat{\beta})} G_1(\vec{n},\vec{m})
\]
\[
  +
  B^{-1}\frac{J_2(\hat{2\beta})}{I_2(2\hat{\beta})}\hat{\ell}
  G_2(\vec{n},\vec{m})]
  C^{(s)}_{n,m;\alpha,\beta}(\vec{x};t_1,t_2)\}
  +\frac{da}{a}\{\frac{\Lambda^{1-d}}{2N}\frac{A^{-1}B^{-1}}{N^2}
  \sum_n\sum_m\sum_{\alpha}\int d^dx\int dt
\]
\[
  [\frac{J_2(\hat{\beta})}{2I_2(\hat{\beta})}K_3(\vec{n},\vec{m})
  +\frac{J_1(\hat{\beta})}{I_1(\hat{\beta})} K_2(\vec{n},\vec{m})
  +\frac{B}{A}\frac{J_1(\hat{\beta})}{I_1(\hat{\beta})} K_3(\vec{n},\vec{m})]
  O^{(s)}_{n,m;\alpha}(\vec{x},t)\}
\]
\begin{equation}
\label{eq-048}
  +\frac{da}{a}\{\frac{\Lambda^{1-d}}{2N}\frac{A^{-1}B^{-1}}{N^2}
  \frac{J_1(\hat{\beta})}{I_1(\hat{\beta})}
  \sum_n\sum_m\sum_{\alpha}\int d^dx\int dt
  F(\vec{n},\vec{m})Q^{(B)}_{n,m;\alpha}(\vec{x},t)\}
\end{equation}

\noindent In Eq.\ref{eq-048} the time integration introduces:

\begin{equation} 
\label{eq-049}
  J_1(\hat{\beta})\simeq I_1(\hat{\beta}),
	\;\;\;\;\;
  J_2(\hat{\beta})\simeq I_2(\hat{\beta}).
\end{equation}

\noindent The integral in Eq.\ref{eq-049} depends explicitly on the
dimensionless $\hat{\beta}$. At the scale $b=1$ we have
$\hat{\beta}(b=1)=\hat{\beta}\gg 1$ and for $b=b_T=\beta/a$ we have
$\hat{\beta}(b)=1$ and have to stop scaling. By Using the OPE
rules we generate the polynomials $G_1$, $G_2$, $K_1$, $K_2$, $K_3$, and $F$.
These polynomials are obtained from the microscopic couplings and the OPE results
obtained at second order ( the polynomials $R$, $S$, $T$, $L$, and $M$). 

\begin{equation}  
\label{eq-050}
  G_1(\vec{n},\vec{m})=-d^{(s)}_3[\frac{3}{2}S(\vec{n},\vec{m})
  +2R(0)\delta_{n,m} -T(\vec{n},\vec{m})],
\end{equation}

\begin{equation}   
\label{eq-051}
  G_2(\vec{n},\vec{m})=-d^{(s)}_3[2\hat{M}(0)e^{(s)}_3(0)\delta_{n,m}
  +\frac{3}{2}\hat{M}(\vec{n},\vec{m})e^{(s)}_2(\vec{n},\vec{m})-2
  \hat{M}(\vec{n},\vec{m})e^{(c)}_2(\vec{n},\vec{m})],
\end{equation}

\begin{equation}    
\label{eq-052}
  K_1(\vec{n},\vec{m})=-d^{(s)}_3[2\sum_{\vec{l}}R(\vec{l},\vec{m})+
  \frac{1}{2}S(\vec{n},\vec{m})-T(\vec{n},\vec{m})],
\end{equation}

\begin{equation}     
\label{eq-053}
  K_2(\vec{n},\vec{m})=-d^{(s)}_3[2\sum_{\vec{l}}\vec{L}(\vec{n},\vec{l})
  e^{(s)}_3(\vec{l},\vec{m})+\frac{1}{4}\vec{L}(\vec{n},\vec{m})
  e^{(s)}_2(0)],
\end{equation}

\begin{equation}      
\label{eq-054}
  K_3(\vec{n},\vec{m})=-d^{(s)}_3[2\sum_{\vec{l}}\vec{M}(\vec{n},\vec{l}) 
  e^{(s)}_3(\vec{l},\vec{m})+\frac{1}{2}\vec{M}(\vec{n},\vec{m}) 
  e^{(s)}_2(\vec{n},\vec{m})+2\vec{M}(\vec{n},\vec{m})
  e^{(c)}_2(\vec{n},\vec{m})], 
\end{equation}

\begin{equation}       
\label{eq-055}
  F(\vec{n},\vec{m})=-d^{(s)}_3[2\vec{L}(\vec{n},\vec{m})
  e^{(s)}_3(\vec{n},\vec{m})-\frac{3}{4}\delta_{n,m}
  e^{(s)}_2(0)\hat{L}(\vec{n},\vec{m})].
\end{equation}

\noindent Next we compute:

\[
  \langle\tilde{S}_{int}\tilde{S}^2_D\rangle_{da}=\frac{da}{a}\{
  \frac{\Lambda^{2-d}}{N}\frac{A^{-1}B^{-1}}{N^2}\sum_n\sum_m\sum_{\alpha}
  \sum_{\beta} \int d^dx\int dt_1\int dt_2 [
  \frac{J_1(\hat{\beta})}{2I_1(\hat{\beta})}(d^{(s)}_3)^2
  \hat{M}(\vec{n},\vec{m})(4N-1)
\]
\begin{equation}
\label{eq-056}
  +\frac{J_2(2\hat{\beta})}{I_2(2\hat{\beta})}(d^{(s)}_3)^2
  (2\hat{L}(0)\delta_{n,m}+4\sum_{\hat{l}}\hat{M}(\vec{l},\vec{m})
  -\hat{M}(\vec{n},\vec{m}))]
  C^{(s)}_{n,m;\alpha,\beta}(\vec{x};t_1,t_2)\}.
\end{equation}

\noindent The OPE in Eq.\ref{eq-056} determines the behavior of the 
Cooperon as a function of the polynomials
$\hat{M}$ (see Eq.\ref{eq-047}) and Cooperon coupling $d^{(s)}_3$.

\[
  \frac{1}{3!} \langle\tilde{S}^3_{int}\rangle_{da}=\frac{da}{a}\{
  \frac{\Lambda^{1-d}}{2N}\frac{(A^{-1})^2}{3!4N^2}
  \frac{J_1(\hat{\beta})}{I_1(\hat{\beta})} 
  \sum_n\sum_m\sum_{\alpha}\int d^dx\int dt
\]
\begin{equation} 
\label{eq-057}
  [W(\vec{n},\vec{m})Q^{(B)}_{n,m;\alpha}(\vec{x},t)-
  V(\vec{n},\vec{m})H^{(B)}_{n,m;\alpha}(\vec{x},t)+
  U(\vec{n},\vec{m})O^{(s)}_{n,m;\alpha}(\vec{x},t)]\}
\end{equation}

\noindent where the functions $U$, $V$, and $W$ are defined in terms of the
microscopic couplings and the second order functions $R$ and $S$ defined
in Eq.\ref{eq-039}.

\[
  W(\vec{n},\vec{m})=-[2R(\vec{n},\vec{m})e^{(s)}_3(\vec{n},\vec{m})
  +\frac{3}{4}\delta_{n,m}(R(0)e^{(s)}_2(0)+S(0)e^{(s)}_3(0))]
\]
\[
  V(\vec{n},\vec{m})=4S(\vec{n},\vec{m})e^{(s)}_2(\vec{n},\vec{m})
\]
\begin{equation}  
\label{eq-058}
  U(\vec{n},\vec{m})=2\sum_{\vec{l}}R(\vec{n},\vec{l})
  e^{(s)}_3(\vec{l},\vec{m})+\frac{1}{4}(
  R(\vec{n},\vec{m})e^{(s)}_2(0)+S(0)e^{(s)}_3(\vec{n},\vec{m}))
\end{equation}

\noindent and

\[
  \frac{1}{3!} \langle\tilde{S}^3_D\rangle_{da}=\frac{da}{a}\{
  -\frac{\Lambda^{2-d}}{N}(B^{-1})^2\frac{J_2(2\hat{\beta})}{I_2(2\hat{\beta})}
  \frac{16}{3!}\sum_n\sum_m\sum_{\alpha}\sum_{\beta}\int d^dx\int
  dt_1\int dt_2
\]
\begin{equation}   
\label{eq-059}
  (d^{(s)}_3)^3(1-\frac{1}{4N})^3C^{(s)}_{n,m;\alpha,\beta}(\vec{x};t_1,t_2)\}
\end{equation}

\noindent Using the results given in Eqs.\ref{eq-029}-\ref{eq-059} we will
obtain the RG equations.\\

\section{The RG equations in the Quantum limit}
\label{sec-5}

The quantum region is defined by $\Lambda/b_T<|q|<\Lambda$ where
$b_T=\frac{v_F\Lambda}{T}$. In principle it is possible that before the
scale $b_T$ has been reached, one of the coupling constants has reached
values of order one. If this happens at a scale $b_o<b_T$ we have to stop at
$b_o$ and for the interval $\frac{\Lambda}{b_T}\le |q|<\frac{\Lambda}{b_o}$ we
have a different theory. If the Cooperon coupling constant
$\frac{\hat{t}}{N}\propto(k_F\ell)^{-1}$ reaches values of order one at
$b_o<b_T$ we must crossover to the Finkelstein diffusion theory. From the
other hand if one of the two-body interactions reaches large values we have to
construct a new theory. If the two-body interaction which grows under scaling
is the Cooper coupling constant we have to construct a theory based on a
superconductivity with disorder. We will consider here the situation where
the effects of interactions are such that the value of $b_o\equiv b_{Dif}$
obeys $b_{Dif}>b_T$ or $b_o\equiv b_{SC}$, $b_{SC}<b_T$ ($b_{SC}$ is the
length scale where the Cooper coupling constant diverges.). Therefore we
will ignore the diffusive region.\\

We introduce the following rescaled coupling constants:

\[
  d^{(s)}_3=\hat{t}B; \;\;\;\;\;
  e^{(c)}_2(\vec{n},\vec{m})=\hat{e}^{(c)}_2(\vec{n},\vec{m})A;
\]
\[
  \Gamma^{(c)}_2(\vec{n},\vec{m})=\hat{\gamma}^{(c)}_2(\vec{n},\vec{m})A;
  \;\;\;\;\;
  e^{(s)}_2(\vec{n},\vec{m})=\hat{e}^{(s)}_2(\vec{n},\vec{m})A;
\]
\[
  \Gamma^{(s)}_2(\vec{n},\vec{m})=\hat{\gamma}^{(s)}_2(\vec{n},\vec{m})A; 
  \;\;\;\;\;
  e^{(s)}_3(\vec{n},\vec{m})=\hat{e}^{(s)}_3(\vec{n},\vec{m})A;
\]
\begin{equation}
\label{eq-060}
  \Gamma^{(s)}_3(\vec{n},\vec{m})=\hat{\gamma}^{(s)}_3(\vec{n},\vec{m})A.
\end{equation}

\noindent where the constants $A$ and $B$ are defined in Eqs.\ref{eq-040} and
\ref{eq-044}. In the quantum regime the number of channels obeys
$N_0\rightarrow N(b)=\pi(\frac{k_F}{\Lambda/b})=N_0b$.
Due to the fact that when the cutoff $\Lambda$ is reduced to $\Lambda/b$
the number of channels scales like $N(b)=N_0b$, it follows that the naive
scaling dimension of the interaction and disorder will be

\[
  \frac{\gamma}{N}\stackrel{\Lambda/b}{\longrightarrow}
  \frac{\gamma}{N}b^{2-d}  
  \;\;\;\;
  and
  \;\;\;\;
  \frac{\hat{t}}{N}\stackrel{\Lambda/b}{\longrightarrow}
  \frac{\hat{t}}{N}b^{3-d}.
\]

\noindent We observe that the interaction becomes marginal while the
disorder is relevant. In the opposite situation where the number of
channels does not scale, we have: $\gamma\rightarrow\gamma b^{1-d}$
and $\hat{t}\rightarrow\hat{t}b^{2-d}$.\\

For the disorder Cooperon coupling constant $\hat{t}$ we
have the scaling equation:

\[
  \frac{d\hat{t}}{d\ln b}=\hat{t}[1-\frac{1}{N}(\frac{3}{4}
  \hat{\gamma}^{(s)}_2(\vec{n},\vec{m})-\hat{\gamma}^{(c)}_2(\vec{n},\vec{m})
  +\frac{J_1(\hat{\beta})}{I_1(\hat{\beta})}\delta_{n,m}
  [\hat{\gamma}^{(s)}_3]^2_{n,m})]
\]
\begin{equation}
\label{eq-061}
  +2\hat{t}^2[1-\frac{1}{N}\frac{J_1(\hat{\beta})}{I_1(\hat{\beta})}
  (\frac{3}{2}\hat{\gamma}^{(s)}_2(\vec{n},\vec{m})-2
  \hat{\gamma}^{(c)}_2(\vec{n},\vec{m}))-\frac{1}{N}
  \frac{J_2(\hat{\beta})}{I_2(\hat{\beta})}(
  3\langle\hat{\gamma}^{(s)}_2\rangle
  -2\langle\hat{\gamma}^{(c)}_2\rangle)].
\end{equation}

\noindent In Eq.\ref{eq-061} we use the notation:

\begin{equation}
\label{eq-062}
  [\hat{\gamma}^{(s)}_3]^2_{n,m} \equiv \frac{1}{N} \sum_{\vec{l}}
  \hat{\gamma}^{(s)}_3(\vec{n},\vec{l})\hat{\gamma}^{(s)}_3(\vec{l},\vec{m})
\end{equation}

\begin{equation} 
\label{eq-063}
  \langle\hat{\gamma}^{(s)}_2\rangle=\frac{1}{N}\sum_{\vec{l}}
  \gamma^{(s)}_2(\vec{l},\vec{n}),
  \;\;\;\;\;
  \langle\hat{\gamma}^{(c)}_2\rangle=\frac{1}{N}\sum_{\vec{l}}
  \gamma^{(c)}_2(\vec{l},\vec{n}).
\end{equation}

\noindent From Eq.\ref{eq-061} we see that we can have a M-I transition in
two dimensions when the p-p interaction $\gamma^{(s)}_3$ and the p-h
$\gamma^{(s)}_2$ increases such that the linear term in ``$\hat{t}$"
becomes negative (see Eq.\ref{eq-061}). We observe in Eq.\ref{eq-061}
that the
effect of the p-h singlet $\gamma^{(c)}_2$ is opposite to the p-h triplet
$\gamma^{(s)}_2$. $\gamma^{(c)}_2$ enhances the localization while
$\gamma^{(s)}_2$ drives the system metallic. This is consistent with the known
fact that a ``Hartree" term ($\gamma^{(c)}_2$) favors localization while the
``Fock" exchange term ($\gamma^{(s)}_2$) drives the system metallic. From
dimensional analysis it follows that Eq.\ref{eq-061} must be linear in
$\hat{t}$. In addition we have that the number of channels obey the scaling
law, $N=N(b)=N_ob$.\\

The scaling equation for the particle-hole singlet is:

\[
  \frac{d\hat{\gamma}^{(c)}_2(\vec{n},\vec{m})}{d \ln b}=\frac{1}{N}\{
  (\hat{\gamma}^{(s)}_3(\vec{n},\vec{m}))^2+
  \frac{J_1(\hat{\beta})}{I_1(\hat{\beta})} [
  \hat{t}\langle\hat{\gamma}^{(s)}_3\rangle(
  \hat{\gamma}^{(s)}_3(\vec{n},\vec{m})(1-3\delta_{n,m})+
  \frac{3}{2}\delta_{n,m}\hat{\gamma}^{(s)}_2(0))
\]
\begin{equation} 
\label{eq-064}
  -\frac{1}{24}[\hat{\gamma}^{(s)}_3]^2_{n,m}(
  \hat{\gamma}^{(s)}_3(\vec{n},\vec{m})(4-3\delta_{n,m})
  +\frac{3}{2}\delta_{n,m}\hat{\gamma}^{(s)}_3(0))]\}
  +\frac{1}{2}\delta_{n,m}
  \frac{d\hat{\gamma}^{(s)}_3(\vec{n},\vec{m})}{d \ln b}
\end{equation}

\noindent and the particle-hole triplet 
$ \hat{\gamma}^{(s)}_2(\vec{n},\vec{m})$ is given by:

\begin{equation}  
\label{eq-065}
  \frac{d\hat{\gamma}^{(s)}_2(\vec{n},\vec{m})}{d \ln b}=\frac{1}{N}\{
  -(\hat{\gamma}^{(s)}_2(\vec{n},\vec{m}))^2+\frac{1}{6N}
  \frac{J_1(\hat{\beta})}{I_1(\hat{\beta})}
  (\hat{\gamma}^{(s)}_2(\vec{n},\vec{m}))^3\})+
  2\delta_{n,m}\frac{d\hat{\gamma}^{(s)}_3(\vec{n},\vec{m})}{d \ln b}
\end{equation}

\noindent From Eqs.\ref{eq-064} and \ref{eq-065} we see that the
particle-particle channel affects the particle-hole singlet. In addition for
$\vec{n}=\vec{m}$ the particle-particle channel 
$\hat{\gamma}^{(s)}_3(\vec{n},\vec{n})$ is identicle to the particle-hole
triplet $\frac{1}{2}\hat{\gamma}^{(s)}_2(\vec{n},\vec{n})$,
$\hat{\gamma}^{(s)}_3(\vec{n},\vec{n})=
\frac{1}{2}\hat{\gamma}^{(s)}_2(\vec{n},\vec{n})$.\\

The particle-particle singlet term obeys the scaling equation:

\[
  \frac{d\hat{\gamma}^{(s)}_3(\vec{n},\vec{m})}{d \ln b}=
  -\frac{1}{2}[\hat{\gamma}^{(s)}_3]^2_{n,m}
  +\hat{t}\langle\hat{\gamma}^{(s)}_3\rangle+\frac{1}{3!}
  \frac{J_1(\hat{\beta})}{I_1(\hat{\beta})}[\hat{\gamma}^{(s)}_3]^3_{n,m}
\]
\[
  -2\frac{J_2(\hat{\beta})}{I_2(\hat{\beta})}\hat{t}
  \langle(\hat{\gamma}^{(s)}_3)^2\rangle
  -4\frac{J_1(\hat{\beta})}{I_1(\hat{\beta})}\hat{t}
  \langle(\hat{\gamma}^{(s)}_3)^2\rangle
  +8\frac{J_2(\hat{\beta})}{I_2(\hat{\beta})}\hat{t}^2
  \langle(\hat{\gamma}^{(s)}_3)\rangle
  +8\frac{J_1(\hat{\beta})}{I_1(\hat{\beta})}\hat{t}^2
  \langle(\hat{\gamma}^{(s)}_3)\rangle
\]
\[
  +\frac{1}{N}\{[\hat{t}+4\hat{t}^2(
  \frac{J_1(\hat{\beta})}{I_1(\hat{\beta})}+\frac{A}{B}
  \frac{J_2(\hat{\beta})}{I_2(\hat{\beta})})]
  [\frac{1}{2}\hat{\gamma}^{(s)}_2(\vec{n},\vec{m})
  -2\hat{\gamma}^{(c)}_2(\vec{n},\vec{m})]\}
  +\frac{1}{N} \{\frac{1}{8}\hat{\gamma}^{(s)}_3(\vec{n},\vec{m})
  \hat{\gamma}^{(s)}_3(0)-\frac{1}{16}\hat{\gamma}^{(s)}_3(\vec{n},\vec{m})
  \hat{\gamma}^{(s)}_2(0)
\]
\begin{equation}   
\label{eq-066}
  -\frac{1}{3!}\frac{J_1(\hat{\beta})}{I_1(\hat{\beta})}\delta_{n,m}
  [\hat{\gamma}^{(s)}_3]^2_{n,m}\hat{\gamma}^{(s)}_3(0)
  +\frac{1}{2!3!}\frac{J_1(\hat{\beta})}{I_1(\hat{\beta})}
  [\hat{\gamma}^{(s)}_3]^2_{n,m}\hat{\gamma}^{(s)}_2(0)
  -\frac{J_2(2\hat{\beta})}{I_2(2\hat{\beta})}
  \langle(\hat{\gamma}^{(s)}_3)\rangle
  (\frac{1}{2}\hat{\gamma}^{(s)}_2(0)-\hat{\gamma}^{(s)}_3(0)\delta_{n,m})\}
\end{equation}

\noindent where

\[
  [\hat{\gamma}^{(s)}_3]^2_{n,m}=\frac{1}{N}\sum_{\vec{l}}
  \hat{\gamma}^{(s)}_3(\vec{n},\vec{l})\hat{\gamma}^{(s)}_3(\vec{l},\vec{m})
\]
\[
  \langle(\hat{\gamma}^{(s)}_3)^2\rangle=\frac{1}{N}\sum_{\vec{l}}
  (\hat{\gamma}^{(s)}_3(\vec{n},\vec{l}))^2
\]
\[ 
  \langle(\hat{\gamma}^{(s)}_3)\rangle=\frac{1}{N}\sum_{\vec{l}} 
  \hat{\gamma}^{(s)}_3(\vec{n},\vec{l})
\] 
\begin{equation}    
\label{eq-067}
  [\hat{\gamma}^{(s)}_3]^3_{n,m}=\frac{1}{N^2}\sum_{\vec{l}}
  \sum_{\vec{l}^{\prime}}\hat{\gamma}^{(s)}_3(\vec{n},\vec{l})
  \hat{\gamma}^{(s)}_3(\vec{l},\vec{l}^{\prime})
  \hat{\gamma}^{(s)}_3(\vec{l}^{\prime},\vec{m}).
\end{equation}

\noindent The scaling relation for the forward part are trivial:

\begin{equation}     
\label{eq-068}
  \frac{d\Gamma^{(c)}}{d\ln b}= \frac{d\Gamma^{(s)}}{d\ln b}=0
\end{equation}

\noindent The set of Eqs.\ref{eq-064}-\ref{eq-066} show that in the limit of
$N\rightarrow\infty$ the interaction is controlled only by the
particle-particle singlet $\hat{\gamma}^{(s)}_3(\vec{n},\vec{m})$. In
addition we observe that the disorder renormalizes the 
$\hat{\gamma}^{(s)}_3(\vec{n},\vec{m})$. We observe that the scaling equation
for $\hat{\gamma}^{(s)}_3(\vec{n},\vec{m})$ can be negative at $b=1$.
The origin of the negative
term is given by Eq.\ref{eq-037}, where it has been shown that at short times
the Cooperon behaves like a Cooper p-p singlet. As a result the initial
values of the particle-particle singlet
$\hat{\gamma}_3^{(s)}(\vec{n},\vec{m}; b=1)$ are replaced by
$\hat{\gamma}_3^{(s)}(\vec{n},\vec{m}; b=1)-2v_F(\frac{B}{A})\hat{t}$.
In Eqs.\ref{eq-061},
\ref{eq-064}, \ref{eq-065}, and \ref{eq-066} the scaling of the number of the
channels is stopped when diffusive region is reached. At finite temperature we
stop scaling at the scale $b=b_T=E_F/T$. This will fix the
number of channels to $\bar{N}\equiv N_T=E_F/T$ (see ref.\cite{09}).
It might be possible that in two dimensions the decoherency introduced by the
temperature might be stronger than $T$. This might be the case if we have in
mind dephasing effects in two dimensions which can define an effective
temperature $T_{eff}(T)>T$ replacing $\bar{N}$ by $E_F/T_{eff}$.\\

\section{The conducting phase due to the superconducting instability in the
quantum region}
\label{sec-6}

In the low temperature limit we can ignore all the many body effect except
the particle-particle singlet $\hat{\gamma}^{(s)}_3(\vec{n},\vec{m})$. The
reason being the $1/N$ factor which appears in Eqs.\ref{eq-064} and
\ref{eq-065} and is missing for the particle-particle singlet in
Eq.\ref{eq-066}. The growth of the number of
channels $N(b)$ is determined by the topology of the Fermi surface. In
particular this is the case for spherical Fermi surface where $N(b)=N_ob$.
For $T\not= 0$ we obtain $N(b=b_T)=\frac{E_F}{T}$. (In chapter
\ref{sec-8} we will consider non-spherical Fermi surface with repulsive
interaction which might lead to a Ferromagnetic instability.)\\

Due to the fact that the $1/N$ factor is only absent for the particle-particle
singlet, we will investigate the
problem in the parameter space $(\hat{\gamma}^{(s)}_3,\hat{t})$
using the angular momentum representation:

\[
  \gamma^{(s)}_3(r)\equiv\gamma_r=\int_0^{\pi}\frac{d\theta}{\pi}
  \gamma^{(s)}_3(\theta)\cos(r\theta),
  \;\;\;\;\;
  r=0,2,4,\cdots.
\]

\noindent For the singlet case $r=0$ we have $\gamma_{r=0}=\gamma_o$:

\begin{equation}
\label{eq-069}
  \gamma_o\simeq\frac{1}{N}\sum_{\vec{l}}
  \hat{\gamma}^{(s)}_3(\vec{l},\vec{n})
\end{equation}

\noindent From Eq.\ref{eq-066} we obtain to leading order in $1/N$
the following equation for particle-particle singlet:

\begin{equation}
\label{eq-070}
  \frac{d\gamma_o}{d\ln b}=-\frac{1}{2}\gamma_o^2
  +\gamma_o\hat{t}-6\gamma_o^2\hat{t}+16\gamma_o\hat{t}^2
  +\frac{1}{3!}\gamma_o^3,
\end{equation}

\noindent with

\[
  \gamma_o(b=1)\rightarrow\gamma_o(b=1)-2\tilde{v_F}\hat{t}
\]

\noindent In Eq.\ref{eq-070} we have used $I_1\sim I_2\sim J_1\sim 1$ and
$\tilde{v_F}=v_FB/A$ where the constants $A$ and $B$ have been defined
in Eqs.\ref{eq-040} and \ref{eq-044}.\\

\noindent We investigate Eq.\ref{eq-070} in the limit of weak disorder
$\hat{t}\rightarrow 0$. We find that even for positive value of $\gamma_o$
the effect of disorder is to drive $\gamma_o(b)$ to negative values.
The reason for this is the fact that the negative linear term in $\hat{t}$ can
cause an initial negative value for $\gamma_o(b=1)$. As a result the term
$-\frac{1}{2}\gamma_o^2$ (for negative value of $\gamma_o$, $\gamma_o(b=1)<0$)
might drive the particle-particle interaction towards a superconducting
instability. This behavior can be seen in the following way. In the
limit of $\hat{t}\rightarrow 0$ we keep in Eq.\ref{eq-070} only the
two first order terms and obtain the solution for $\gamma_o(b)$:

\begin{equation}
\label{eq-071}
  \gamma_o(b=e^{\ell})=\gamma_o(b=1)\; exp(\int_0^{\ell}\hat{t}(x)\;dx)
[1+\frac{1}{2}\gamma_o(b=1)\int_0^{\ell}dy\;exp(\int_0^{y}\hat{t}(x)\;dx)]^{-1} 
\end{equation}

\noindent For $\gamma_o(b=1)<0$, $\gamma_o(b)$ diverges at a length scale
$b=b_{SC}\equiv\frac{v_F\Lambda}{T_{SC}}$ where $T_{SC}$ represents
the superconducting instability temperature,

\[
  T_{SC}=v_F\Lambda(1+\frac{2\hat{t}}{|\gamma_o(b=1)|})^{-\frac{1}{t}}
  \stackrel{\hat{t}\rightarrow 0}{\longrightarrow}
  v_F\Lambda\; exp(-\frac{2}{|\gamma_o(b=1)|}).
\]

\noindent Next we consider the RG equation for the Cooperon (see
Eq.\ref{eq-061} with
$J_1(\hat{\beta})\sim I_1(\hat{\beta})\sim I_2(\hat{\beta})\sim 1$).
From Eq.\ref{eq-061} we observe that in the limit of
vanishing interactions the Cooperon
coupling constant scales like $\frac{\hat{t}(b)}{N(b)}\sim
\frac{\hat{t}(b=1)}{N_o}[1-2\hat{t}(b=1)\log b]^{-1}$ and diverges at
$b\equiv b_{Loc} \equiv \frac{v_F\Lambda}{T_{Loc}}$
($b_{Loc}\ge b_{Dif}$, $T_{Dif}\ge T_{Loc}$),
$T_{Loc}\simeq v_F\Lambda\;exp[-\frac{1}{2\hat{t}(b=1)}]$.
In order to understand the physics of the system we have to compare
the physical temperature $T$ with the other two, $T_{SC}$ and $T_{Loc}$.
We have to consider separately the cases:
a) $T<T_{SC}<T_{Loc}$; b) $T<T_{Loc}<T_{SC}$; c) $T_{SC}<T_{Loc}<T$;
d) $T_{SC}<T<T_{Loc}$; e) $T_{Loc}<T<T_{SC}$; f) $T_{Loc}<T_{SC}<T$.\\

\noindent a) $T<T_{SC}<T_{Loc}$\\
This is the localized case where the mean free path ``$\ell$" is the shortest
length scale in the problem. This case will not be analyzed here. Most of the
work in the past has been concentrated towards this case, in particular the
Finkelstein theory which has investigated the interactions within the diffusion theory.\\

\noindent b) $T<T_{Loc}<T_{SC}$\\
Here the shortest length scale is the Cooper coherence length. Physically
one can describe this region by a system of disorder bosons (the bosons
describe the pairs). The critical theory might correspond to a disorder X-Y
model.\\

\noindent c) $T_{SC}<T_{Loc}<T$\\
This is a region where interactions are not important. The physics is controlled
by classical hopping transport.\\

\noindent d) $T_{SC}<T<T_{Loc}$\\
As in case a) here the system is localized. This case will not be considered
here. (See the Finkelstein theory.)\\

\noindent e) $T_{Loc}<T<T_{SC}$\\
Again a bosonic X-Y theory with disorder is applicable here as in case b).\\

\noindent f) $T_{Loc}<T_{SC}<T$\\
In this region we will have transport controlled by pair breaking.\\

In the rest part of this section we will investigate the RG equation for the
negative particle-particle singlet $\hat{\gamma}_3^{(s)}(\vec{n},\vec{m})$
and the Cooperon coupling constant $\hat{t}$. In agreement with
Eq.\ref{eq-069} we introduce the angular momentum representation for the
Cooper and Cooperon channels:
$\gamma_o=\frac{1}{N}\sum_{\vec{\l}}\hat{\gamma}_3^{(s)}(\vec{l},\vec{n})$,
$t_o=\frac{1}{N}\sum_{\vec{\l}}\hat{t}(\vec{l},\vec{n})$. We obtain from
Eq.\ref{eq-061} and Eq.\ref{eq-070} the following RG equations:

\[
  \frac{d\lambda}{d\ln b}=\frac{1}{2}\lambda^2+\lambda t_o,
  \;\;\;\;\;
  \lambda\equiv -\gamma_o
\]
\[
  \frac{dt_o}{d\ln b}=t_o[1-(\frac{\lambda}{N})^2]+2t_o^2
\]
\begin{equation}
\label{eq-072}
  \rho(b)\propto\frac{t_o(b)}{N(b)}\equiv\bar{t}_o(b),
  \;\;\;\;\;
  N(b)=N_ob
\end{equation}

\noindent $\rho(b)$ is the resistance with $b$ restricted to
$1<b\le\frac{v_F\Lambda}{T}$. From Eq.\ref{eq-072} we observe that in the 
limit
$b\rightarrow\infty$ ($T\rightarrow 0$) the parameter $\lambda$ diverges.
In particular we observe that the ratio
$\frac{\lambda(b)}{N(b)}\stackrel{b\rightarrow\infty}{\longrightarrow}\infty$.
As a result the RG equation behaves like
$\frac{dt_o}{d\ln b}=-t_o(\frac{\lambda}{N})^2$. Due to the large value of
$(\frac{\lambda}{N})^2$ it follows that
$t_o(b)\stackrel{b\rightarrow\infty}{\longrightarrow}0$. As a result we obtain
a superconducting ground state. At finite temperature we consider the case
$b_T<b_{SC}<b_{Loc}$. We substitute the solution of $\lambda(b)$ into
$t_o(b)$ and obtain

\begin{equation}
\label{eq-073}
  \bar{t}_o(b_T)=\frac{t_o}{N_o}
  exp\{-\int_0^{\log b_T}(\frac{\lambda(x)}{N(x)})^2 dx\}
  \sim \frac{t_o}{N_o}
  exp\{-\frac{(4/N_o^2)T_{SC}}{|T-T_{SC}|}\},
  \;\;\;
  T>T_{SC}
\end{equation}

\noindent $N_o\simeq\frac{\pi k_F}{\Lambda}\sim 1$ and $T_{SC}$ is given by
Eq.\ref{eq-071}. As a result we obtain that the resistance obeys
$\rho(T)\stackrel{T\rightarrow T_{SC}}{\longrightarrow}0$, 
$  \rho(T)\sim\;Const.\; exp\{-\frac{(4/N_o^2)T_{SC}}{|T-T_{SC}|}\}$.
To conclude this section ($\gamma_o<0$) we remark that the transport
data \cite{03} show some similarity with the one reported for disorder
bosons in ref.\cite{04}. This might suggest that the correct starting point
might be a disordered bosonic system instead of a diffusion theory \cite{02}.\\

\section{The RG equation at finite temperature}
\label{sec-7}

At a temperature $T$ the scaling is restricted to
$\frac{\Lambda}{b_T}<|q|<\Lambda$ where $b_T=\frac{v_F\Lambda}{T}$.
In this interval the number of channels is restricted to
$\bar{N}=N(b_T)=\frac{E_F}{T}$,
with $N(b)$ obeying the condition $N_o<N(b)\le\bar{N}$. We replace in
Eqs.\ref{eq-061}-\ref{eq-068} $J_1(\hat{\beta})\sim J_2(\hat{\beta})\sim
I_1(\hat{\beta})\sim I_2(\hat{\beta})\sim 1$ and find a simplified form

\[
  \frac{d\hat{t}}{d\ln b}=\epsilon(b)\hat{t}-\frac{\hat{t}}{N}(\frac{3}{4}
  \hat{\gamma}^{(s)}_2(\vec{n},\vec{m})-\hat{\gamma}^{(c)}_2(\vec{n},\vec{m})
  +\delta_{n,m}[\hat{\gamma}^{(s)}_3]^2_{n,m})
\]
\begin{equation}
\label{eq-074}
  +2\hat{t}^2[1-\frac{1}{N}(\frac{3}{2}
  \hat{\gamma}^{(s)}_2(\vec{n},\vec{m})-2\hat{\gamma}^{(c)}_2(\vec{n},\vec{m}))
  -\frac{1}{N}(3\langle\hat{\gamma}^{(s)}_2\rangle
  -2\langle\hat{\gamma}^{(c)}_2\rangle)]
\end{equation}

\noindent The parameter $\epsilon(b)$ controls the crossover at finite
temperatures. $\epsilon(b)$ is given by, $\epsilon(b)=1$ for $b<b_T$ and
$\epsilon(b)\simeq 0$ for $b>b_T$. Eq.\ref{eq-074} replaces the scaling
Eq.\ref{eq-061} for the disorder
coupling constant $\hat{t}$. In Eq.\ref{eq-074} we observe that the
interaction
has produced a shift in the critical dimensionality. The disorder parameter
$\hat{t}$ has accumulated a finite anomalous dimension,
$\frac{1}{N}(\frac{3}{4}\hat{\gamma}_2^{(s)}\cdots)$, which will control the
M-I transition. (In the limit $T\rightarrow 0$, $N\rightarrow\infty$
causing this term to disappear.)\\

At finite temperatures the scaling Eqs.\ref{eq-064} and \ref{eq-065} for
the interactions $\hat{\gamma}^{(s)}_2$ and $\hat{\gamma}^{(c)}_2$ are
the same except that linear terms of the form
$[\epsilon(b)-1]\hat{\gamma}^{(s)}_2$ and
$[\epsilon(b)-1]\hat{\gamma}^{(c)}_2$ are added to the Eqs.\ref{eq-064}
and \ref{eq-065}, respectively.
For the particle-particle singlet $\hat{\gamma}^{(s)}_3$ we have

\[
  \frac{d\hat{\gamma}^{(s)}_3(\vec{n},\vec{m})}{d\ln b}=
  [\epsilon(b)-1]\hat{\gamma}^{(s)}_3(\vec{n},\vec{m})-\frac{1}{2}
  [\hat{\gamma}^{(s)}_3]^2_{n,m}+\hat{t}\langle\hat{\gamma}^{(s)}_3\rangle
  +\frac{1}{3!}[\hat{\gamma}^{(s)}_3]^3_{n,m}-6
  \hat{t}\langle(\hat{\gamma}^{(s)}_3)^2\rangle+16
  \hat{t}\langle\hat{\gamma}^{(s)}_3\rangle
\]
\[
  +\frac{1}{N}\{(\hat{t}+8\hat{t}^2)(\frac{1}{2}
  \hat{\gamma}^{(s)}_2(\vec{n},\vec{m})-2\hat{\gamma}^{(c)}_2(\vec{n},\vec{m}))
  \}+\frac{1}{N}\{\frac{1}{8}\hat{\gamma}^{(s)}_3(\vec{n},\vec{m})
  \hat{\gamma}^{(s)}_3(0)-\frac{1}{16}\hat{\gamma}^{(s)}_3(\vec{n},\vec{m})
  \hat{\gamma}^{(s)}_2(0)
\]
\begin{equation}
\label{eq-075}
  -\frac{1}{3!}\delta_{n,m}[\hat{\gamma}^{(s)}_3]^2_{n,m}
  \hat{\gamma}^{(s)}_3(\vec{n},\vec{m})+\frac{1}{2!3!}
  [\hat{\gamma}^{(s)}_3]^2_{n,m}\hat{\gamma}^{(s)}_2(0)
  -\langle\hat{\gamma}^{(s)}_3\rangle(\frac{1}{2}\hat{\gamma}^{(s)}_2(0)
  -\hat{\gamma}^{(s)}_3(0)\delta_{n,m})\}
\end{equation}

In Eq.\ref{eq-075} we use the same definitions as given in Eq.\ref{eq-067}.
Eq.\ref{eq-075} must be supplemented by the condition
$\frac{1}{2}\hat{\gamma}^{(s)}_2(\vec{n},\vec{n})=
  \hat{\gamma}^{(s)}_3(\vec{n},\vec{n})$ plus Eqs.\ref{eq-064} and
\ref{eq-065}.\\

\section{The scaling equations for the resistivity at finite temperatures
and strong repulsive interactions}
\label{sec-8}

We restrict ourselves to finite temperatures or/and cases where the scaling
of the number of channels is different from $N(b)=N_ob$ (spherical Fermi
surface). For flat Fermi surface the number of the channels does not scale.
We have $N(b)\sim N(b=1)\sim N_o$. At finite temperature for spherical Fermi
surface the number of channels is finite and is restricted by the temperature
$N_o<N(b)<N(b_T)\sim \frac{E_F}{T}$. Since the coupling constants depend on
the number of channels (finite), we will normalize the coupling constant by
$N$, the number of channels

\begin{equation}
\label{eq-076}
  \bar{\gamma}_2^{(c)}\equiv\frac{\hat{\gamma}_2^{(c)}}{N},
  \;\;\;
  \bar{\gamma}_2^{(s)}\equiv\frac{\hat{\gamma}_2^{(s)}}{N},
  \;\;\; 
  \bar{\gamma}_3^{(s)}\equiv\frac{\hat{\gamma}_3^{(s)}}{N},
  \;\;\;
  \bar{t}\equiv\frac{\hat{t}}{N}.
\end{equation}

\noindent As a result the new RG equations are given in terms of the original
Eqs. \ref{eq-075}, \ref{eq-065}, and \ref{eq-066}:

\[
  \frac{d\bar{\gamma}_2^{(c)}}{d\ln b}=\frac{1}{N}
  (\frac{d\hat{\gamma}_2^{(c)}}{d\ln b})-\epsilon_T\bar{\gamma}_2^{(c)};
\]
\[
  \frac{d\bar{\gamma}_2^{(s)}}{d\ln b}=\frac{1}{N} 
  (\frac{d\hat{\gamma}_2^{(s)}}{d\ln b})-\epsilon_T\bar{\gamma}_2^{(s)}; 
\]
\begin{equation} 
\label{eq-077}
  \frac{d\bar{\gamma}_3^{(s)}}{d\ln b}=\frac{1}{N}  
  (\frac{d\hat{\gamma}_3^{(s)}}{d\ln b})-\epsilon_T\bar{\gamma}_3^{(s)}
\end{equation}

\noindent The parameter $\epsilon_T$ depends on the topology of the Fermi
surface and temperature

\begin{equation} 
\label{eq-078}
  \epsilon_T\equiv|\frac{d\ln N(b)}{d \ln b}|,
  \;\;\;
  N_o\le N(b)\le N(b_T).
\end{equation}

\noindent The parameter $\epsilon_T$ takes values of $0\le\epsilon_T\le 1$. The
value of $\epsilon_T=1$ is obtained for spherical Fermi surface $N(b)=N_ob$ and
$\epsilon_T=0$ is obtained for flat Fermi surface or high temperatures,
$N(b)\sim\bar{N}\sim\frac{E_F}{T}$.\\

Here we consider a special case of repulsive interactions such that the
particle-particle singlet and particle-hole triplet are strong in the
backward direction. This
means that the most relevant interactions are those with $\vec{n}=\vec{m}$.
In order to be specific we will consider a special model for which the terms
$\hat{\gamma}_2^{(c)}(\vec{n},\vec{m})$,
$\hat{\gamma}_2^{(s)}(\vec{n},\vec{m})$, and
$\hat{\gamma}_3^{(s)}(\vec{n},\vec{m})$ are zero for $\vec{n}\not=\vec{m}$.
We keep only terms with $\vec{n}=\vec{m}$ and introduce the definition:

\[
  \hat{\gamma}_2^{(c)}\equiv\hat{\gamma}_2^{(c)}(\vec{n},\vec{n}),
  \;\;\;
  \hat{\gamma}_2^{(s)}\equiv\hat{\gamma}_2^{(s)}(\vec{n},\vec{n})
  =2\hat{\gamma}_3^{(s)}(\vec{n},\vec{n})
\]
\begin{equation} 
\label{eq-079}
  \hat{\gamma}_2^{(c)}(\vec{n}\not=\vec{m})\simeq
  \hat{\gamma}_2^{(s)}(\vec{n}\not=\vec{m})\simeq
  \hat{\gamma}_3^{(s)}(\vec{n}\not=\vec{m})\simeq 0
\end{equation}

\noindent Using Eqs. \ref{eq-076}, \ref{eq-077}, and \ref{eq-078} we obtain:

\begin{equation}
\label{eq-080}
  \frac{\bar{\gamma}_2^{(c)}}{d\ln b}=
  -\bar{\gamma}_2^{(c)}(\epsilon_T+\hat{t})
  +\frac{1}{2}\hat{t}\bar{\gamma}_2^{(s)}
  -\frac{3}{4}(\bar{\gamma}_2^{(s)})^2(\hat{t}-\frac{1}{4})
\end{equation}

\begin{equation}
\label{eq-081}
  \frac{\bar{\gamma}_2^{(s)}}{d\ln b}=
  \bar{\gamma}_2^{(s)}(2\hat{t}+8\hat{t}^2-\epsilon_T)
  -(\bar{\gamma}_2^{(s)})^2(\frac{5}{4}+3\hat{t})
  +\frac{5}{24}(\bar{\gamma}_2^{(s)})^3
  -4\bar{\gamma}_2^{(c)}(\hat{t}+8\hat{t}^2)
\end{equation}

\begin{equation} 
\label{eq-082}
  \frac{d\hat{t}}{d\ln b}=
  \hat{t}[1-\frac{3}{4}\bar{\gamma}_2^{(s)}+\bar{\gamma}_2^{(c)}
  -\frac{1}{4}(\bar{\gamma}_2^{(s)})^2]
  +2\hat{t}^2[1-\frac{3}{2}\bar{\gamma}_2^{(s)}+2\bar{\gamma}_2^{(c)}]
\end{equation}

\begin{equation}  
\label{eq-083}
  \bar{t}=\frac{\hat{t}}{N},
  \;\;\;
  N=N(b),
  \;\;\;
  1\le b\le b_T=\frac{E_F}{T}
\end{equation}

\noindent From Eq.\ref{eq-080} we conclude that the particle-hole singlet
$\bar{\gamma}_c$ is irrelevant. Therefore we will take $\bar{\gamma}_c=0$
and ignore Eq.\ref{eq-080}. We will solve the RG equation in the space of
$\bar{\gamma}_2^{(s)}$ and $\hat{t}$ (Eqs. \ref{eq-081} and \ref{eq-082}).
In the parameter space $(\bar{\gamma}_2^{(s)},\hat{t})$ we find a non-trivial
fixed point. In the limit $\epsilon_T\rightarrow 0$ we find
$(\bar{\gamma}_2^{(s)})^*=\frac{8}{5}(\hat{t})^*\simeq\frac{4}{25}$.\\

We linearize the equations around this fixed point and find:
$\hat{t}(b)=\hat{t}^*+(\hat{t}-\hat{t}^*)b^{1/\nu_1}$,
$\nu_1\simeq 1+\frac{2}{25}$ and
$\bar{\gamma}_2^{(s)}(b)=\bar{\gamma}_2^*+(\bar{\gamma}_2^{(s)}-
\bar{\gamma}_2^*)b^{-1/\nu_2}$. 
These equations show that for $\hat{t}<\hat{t}^*$
the disorder decreases and in the same time $\bar{\gamma}_2^{(s)}$ flows to
$\bar{\gamma}_2^*$. For large value of disorder we obtain that $\hat{t}$
increase and $\bar{\gamma}_2^{(s)}$ flows to $\bar{\gamma}_2^*$.
Experimentally the presence of the stable fixed point $\bar{\gamma}_2^*$
might be identified by a power law behavior in the spin-spin
correlation. This is similar to what one has in one dimension and might
corresponds to a spin-liquid phase. For the transport properties, we
believe that our predictions are in a qualitative agreement with the
experiments \cite{01}, we find for the resistivity
$\rho(\bar{t},\bar{\gamma}_2^{(s)},T)$ at a finite temperature and the
dynamical exponent $z\simeq 1$: 
$\rho(\bar{t},\bar{\gamma}_2^{(s)},T)=
\rho(\bar{t}(b),\bar{\gamma}_2^{(s)}(b),Tb^z)=
\rho(\bar{t}^*+(\bar{t}-\bar{t}^*)b^{1/\nu_1}$, $\bar{\gamma}_2^*+
(\bar{\gamma}_2^{(s)}-\bar{\gamma}_2^*) b^{-1/\nu_2},Tb^z)$.
We introduce $Tb^z\simeq T_o\Rightarrow b\simeq(\frac{T_o}{T})^{1/z}$ and use
the definitions $\bar{t}\equiv\frac{\hat{t}}{\bar{N}}$.
As a result we find:

\begin{equation}   
\label{eq-084}
  \rho(\bar{t},\bar{\gamma}_2^{(s)},T)\simeq
  \rho(\bar{t}^*,\bar{\gamma}_2^*,T_o)+{\textstyle const.}(
  \frac{\bar{t}-\bar{t}^*}{\bar{t}^*})(\frac{T_o}{T})^{1/z\nu_1}
\end{equation}

\noindent Eq.\ref{eq-084} shows that for $\bar{t}<\bar{t}^*$ the resistivity
$\rho$ decreases as we lower the temperature and increases when
$\bar{t}>\bar{t}^*$. In order to make contact with the experiments we replace:
$\bar{t}\propto(k_F\ell)^{-1}$, $k_F\propto n_c^{1/2}$,
$\bar{t}^*\propto (n_c^*)^{-1/2}$ ($n_c^*$ is the critical density) and identify$\frac{\bar{t}-\bar{t}^*}{\bar{t}^*}\propto\frac{n_c^*-n_c}{n_c^*}
\equiv\delta$. As a result we find:
$\rho(n_c,\bar{\gamma}_2^{(s)},T)\simeq\rho^*f(\frac{\delta}{T^{1/z\nu_1}})$,
$\rho^*\equiv\rho(n_c^*,\gamma_2^*,T_o)$ which is the result observed in
ref.\cite{03}. We hope that more accurate experiments will confirm the
existence of the suggested fixed point.\\

\section{Conclusion}
\label{sec-9}

A new method for studying many-body systems and disorder has been introduced.
The method is based on the extension of the OPE to two dimensional systems.
Using a real space version of RG we have derived a set of RG equations for
disorder and interaction. We have constructed an alternative theory to the
one constructed by Finkelstein \cite{02}. The basic assumption in
ref.\cite{02} is that the elastic mean free path is the shortest length in
the problem. As a result the multiple elastic scatterings are replaced by
a diffusion theory (the non-linear $\sigma$-model) and the interactions are
considered as a perturbation of the diffusion theory. The method used here is
based on a RG analysis which studies the competitions between the
multiple elastic scattering and the interaction. We identified the following
regions:\\
1) The multiple elastic scattering is the shortest length scale and diverges
first. For this case we agree with the results given in ref.\cite{02} and
do not have anything to add.\\
2) The particle-particle singlet is negative and a superconducting
instability occurs for $T\le T_{SC}$ where $T_{SC}>T_{Loc}$. As a result
one has to treat first the interaction within an effective
Ginzburg-Landau theory. We reproduced a bosonic model (X-Y) which is
perturbed by disorder.\\
3) The interactions are positive and the particle-hole is dominant in the
backward direction. At finite temperature and non-spherical Fermi
surfaces which obey $|\frac{d\ln N(b)}{d\ln b}|\ll 1$ one obtains a
non-trivial fixed point in the plane $(\bar{\gamma}_2^{(s)},\hat{t})$
which separates the conducting from the insulating phase. This fixed
point is characterized by a stable fixed point in the $\bar{\gamma}_2^{(s)}$
direction. No divergence in the particle-hole triplet occurs expect
the infinite correlation length for the spin-spin ferromagnetic correlations
when $\bar{\gamma}_2^{(s)}\rightarrow \bar{\gamma}_2^*$.\\

\begin{center}
  {\bf ACKNOWLEDGMENTS\\}
\end{center} 

D. Schmeltzer would like to thank professor A.M. Finkelstein for his valuable
comment concerning the differences between his theory and the one presented
here.\\

\newpage

\appendix
\section{}
\label{app-1}

The Fermion field $\psi_{\sigma,\alpha}(\vec{x})$ is decomposed into
$N$ Fermions, $\psi_{\sigma,\alpha}(\vec{x})=\sum_{\vec{\omega}=1}^N
e^{ik_F\vec{\omega}\vec{x}}\psi_{\vec{\omega},\sigma,\alpha}$.
Using this representation we
obtain from Eq.\ref{eq-003} the result:

\[
  S_{int}\simeq\int d^dx\int dt\sum_{\sigma,\sigma^{\prime}}\sum_{\alpha}
  \sum_{\vec{\omega}_1}\sum_{\vec{\omega}_2}\sum_{\vec{\omega}_3}
  \sum_{\vec{\omega}_4}
  \delta_{\vec{\omega}_1+\vec{\omega}_2=\vec{\omega}_3+\vec{\omega}_4}
\]
\begin{equation}
\label{eq-a1}
  v(\vec{\omega}_1,\vec{\omega}_2,\vec{\omega}_3,\vec{\omega}_4)
  \psi^{\dagger}_{\vec{\omega}_1,\sigma,\alpha}(\vec{x})
  \psi^{\dagger}_{\vec{\omega}_2,\sigma^{\prime},\alpha}(\vec{x})
  \psi_{\vec{\omega}_3,\sigma^{\prime},\alpha}(\vec{x})
  \psi_{\vec{\omega}_4,\sigma,\alpha}(\vec{x})
\end{equation}

\noindent
where $v(\vec{\omega}_1,\vec{\omega}_2,\vec{\omega}_3,\vec{\omega}_4)$
represents the projection of the screened two-body potential on the Fermi
surface. The presence of the Kroneker-delta function imposes the condition
$\vec{\omega}_1+\vec{\omega}_2=\vec{\omega}_3+\vec{\omega}_4$.
As a result we separate the interaction term into three processes:
1) direct, 2) exchange, and 3) Cooperon channel:

\begin{enumerate}
  \item The direct process is realized when $\vec{\omega}_1=\vec{\omega}_4$,
  $\vec{\omega}_2=\vec{\omega}_3$.
  \item The exchange process:
  $\vec{\omega}_1=\vec{\omega}_3\equiv\vec{\omega}$,
  $\vec{\omega}_2=\vec{\omega}_4\equiv\vec{\omega}^{\prime}$
  \item The Cooperon channel:
  $\vec{\omega}\equiv\vec{\omega}_1=-\vec{\omega}_2$,
  $\vec{\omega}^{\prime}\equiv\vec{\omega}_3=-\vec{\omega}_4$
\end{enumerate}

\noindent As a result Eq.\ref{eq-a1} becomes

\[
  S_{int}\simeq\int d^dx\int dt\sum_{\sigma,\sigma^{\prime}}\sum_{\alpha}
  \sum_{\vec{\omega}}\sum_{\vec{\omega}^{\prime}}\{
  v(0)\psi^{\dagger}_{\vec{\omega},\sigma,\alpha}(\vec{x})
  \psi_{\vec{\omega},\sigma,\alpha}(\vec{x})
  \psi^{\dagger}_{\vec{\omega}^{\prime},\sigma^{\prime},\alpha}(\vec{x})
  \psi_{\vec{\omega}^{\prime},\sigma^{\prime},\alpha}(\vec{x})
\]
\[
  -v(\vec{\omega},\vec{\omega}^{\prime},\vec{\omega},\vec{\omega}^{\prime})
  \psi^{\dagger}_{\vec{\omega},\sigma,\alpha}(\vec{x})
  \psi_{\vec{\omega},\sigma^{\prime},\alpha}(\vec{x})
  \psi^{\dagger}_{\vec{\omega}^{\prime},\sigma^{\prime},\alpha}(\vec{x})
  \psi_{\vec{\omega}^{\prime},\sigma,\alpha}(\vec{x})
\]
\begin{equation}
\label{eq-a2}
  +v(\vec{\omega},-\vec{\omega},\vec{\omega}^{\prime},-\vec{\omega}^{\prime})
  \psi^{\dagger}_{\vec{\omega},\sigma,\alpha}(\vec{x}) 
  \psi_{-\vec{\omega},\sigma^{\prime},\alpha}(\vec{x}) 
  \psi^{\dagger}_{\vec{\omega}^{\prime},\sigma^{\prime},\alpha}(\vec{x}) 
  \psi_{-\vec{\omega}^{\prime},\sigma,\alpha}(\vec{x})\}
\end{equation}

\noindent In Eq.\ref{eq-a2} we observe that the Cooperon channel is identical
to the exchange one if we substitute in the exchange term
$\vec{\omega}^{\prime}=-\vec{\omega}$. This means that we have to take into
consideration this identity in order to avoid double counting.\\

We replace the ``$N$" fermions by $N/2$ pairs of chiral fermions
(see Eq.\ref{eq-005}).
In the second step we replace Eq.\ref{eq-a2} by the current representation:

\[
  S_{int}\simeq\int d^dx\int dt\sum_{\alpha}\sum_{\vec{n},\vec{m}}\{
  (v(0)-\frac{1}{2}v(\vec{n},\vec{m}))
  (J^R_{n,\alpha}(\vec{x})J^R_{m,\alpha}(\vec{x}) 
  +J^L_{n,\alpha}(\vec{x})J^L_{m,\alpha}(\vec{x}))
\]
\[
  -2v(\vec{n},\vec{m})
  (J^R_{n,\alpha}(\vec{x})J^R_{m,\alpha}(\vec{x})  
  +J^L_{n,\alpha}(\vec{x})J^L_{m,\alpha}(\vec{x}))
  +(v(0)-\frac{1}{2}v(\vec{n},\vec{m}+\pi))
  (J^R_{n,\alpha}(\vec{x})J^L_{m,\alpha}(\vec{x})  
\]
\[
  +J^L_{n,\alpha}(\vec{x})J^R_{m,\alpha}(\vec{x})) 
  -2v(\vec{n},\vec{m}+\pi) 
  (J^R_{n,\alpha}(\vec{x})J^L_{m,\alpha}(\vec{x})   
  +J^L_{n,\alpha}(\vec{x})J^R_{m,\alpha}(\vec{x})) 
\]
\[
  +(1-\delta_{n,m})[v(\vec{n},\vec{m})\sum_{\sigma=\uparrow,\downarrow}
  (J^R_{n,\sigma,\alpha;m,\sigma,\alpha}(\vec{x})
  J^L_{n,-\sigma,\alpha;m,-\sigma,\alpha}(\vec{x})
  +J^L_{n,\sigma,\alpha;m,\sigma,\alpha}(\vec{x})
  J^R_{n,-\sigma,\alpha;m,-\sigma,\alpha}(\vec{x}))
\]
\begin{equation}  
\label{eq-a3}
  -v(\vec{n},\vec{m}+\pi)\sum_{\sigma=\uparrow,\downarrow}
  (J^R_{n,\sigma,\alpha;m,-\sigma,\alpha}(\vec{x})
  J^L_{n,-\sigma,\alpha;m,\sigma,\alpha}(\vec{x})
  +J^L_{n,\sigma,\alpha;m,-\sigma,\alpha}(\vec{x}) 
  J^R_{n,-\sigma,\alpha;m,-\sigma,\alpha}(\vec{x}))]\}
\end{equation}

\noindent We introduce the following definitions:

\[
  \tilde{\Gamma}^{(c)}(\vec{n},\vec{m})\equiv v(0)-\frac{1}{2}
  v(\vec{n},\vec{m}),
  \;\;\;\;\;
  \tilde{\Gamma}^{(s)}(\vec{n},\vec{m})\equiv 2v(\vec{n},\vec{m}),
\]
\[
  \Gamma^{(c)}_2(\vec{n},\vec{m})\equiv v(0)-\frac{1}{2}
  v(\vec{n},\vec{m}+\pi),
  \;\;\;\;\;
  \Gamma^{(s)}_2(\vec{n},\vec{m})\equiv 2v(\vec{n},\vec{m}+\pi),
\]
\begin{equation}  
\label{eq-a4}
  \Gamma^{(s)}_3(\vec{n},\vec{m})\equiv\frac{1}{2}
  (v(\vec{n},\vec{m})+v(\vec{n},\vec{m}+\pi)),
  \;\;\;\;\;
  \Gamma^{(t)}_3(\vec{n},\vec{m})\equiv\frac{1}{2}  
  (v(\vec{n},\vec{m})-v(\vec{n},\vec{m}+\pi)).
\end{equation}

\noindent Using the definitions of the interaction operators given in
Eqs.\ref{012} and \ref{eq-013} we obtain the result given in
Eq.\ref{eq-011}.\\

\end{document}